%% file: main.tex
\documentclass[12pt,a4]{article}%
\usepackage{amssymb}
\usepackage{amsfonts}
\usepackage{amsmath}
\usepackage{xfrac}
\usepackage{bm}
\usepackage{bbm}
\usepackage{multirow}
\usepackage{array}
\usepackage{booktabs}
\usepackage{siunitx} 
\usepackage{calc}
\usepackage[nohead]{geometry}
\usepackage{floatpag}
\usepackage[singlespacing]{setspace}
\usepackage[bottom]{footmisc}
\usepackage{indentfirst}
\usepackage{endnotes}
\usepackage{graphicx}%
\usepackage[space]{grffile}
\usepackage{rotating}
\usepackage{pgfplotstable}
\usepackage{adjustbox}
\usepackage[olditem,oldenum]{paralist}
\usepackage{csquotes}
\usepackage{comment} 
\usepackage[firstinits=false, uniquename=false, sorting = nyt, style = apa, language = american, maxcitenames=3, mincitenames = 1]{biblatex}
\DeclareLanguageMapping{american}{american-apa}
\usepackage{hyperref}
\hypersetup{hidelinks}
\usepackage{lscape}
\usepackage[justification=centering]{caption}
\usepackage{threeparttable}
\usepackage{threeparttablex}
\usepackage{longtable}
\usepackage{authblk}
\usepackage{hyphenat}
\usepackage{subfloat}
\usepackage{subcaption}
\usepackage{enumerate}
\usepackage{adjustbox}
\pgfplotsset{compat=1.6}
\makeatletter
\def\@biblabel#1{\hspace*{-\labelsep}}
\makeatother
\usepackage{titlesec}

\title{Trade and pollution: Evidence from India}

\author[1]{Malin Niemi}
\author[2]{Nicklas Nordfors}
\author[3,4]{Anna Tompsett\thanks{Corresponding author: \textit{anna.tompsett@iies.su.se}. We thank Niklas Bengtsson, Tessa Bold, Shon Ferguson, and Guy Michaels for helpful comments and suggestions, along with seminar participants at Stockholm University, the Global Policy Lab at UC Berkeley, and the 13th Annual Giannini Foundation of Agricultural and Resource Economics Student Conference. The authors alone are responsible for any remaining errors. We gratefully acknowledge funding from Jan Wallanders och Tom Hedelius stiftelse samt Tore Browaldhs stiftelse. The authors declare that they have no known competing financial interests or personal relationships that could appear to influence the work reported in this paper.}}

\date{\today}

\affil[1]{Department of Economics, Stockholm University, Sweden}
\affil[2]{Department of Economics, University of Gothenburg, Sweden}
\affil[3]{Beijer Institute of Ecological Economics, Sweden}
\affil[4]{Institute for International Economic Studies, Stockholm University, Sweden}

\begin{document}

\maketitle

\sloppy%

\onehalfspacing

\vspace{-4ex}

\begin{abstract}
\noindent What happens to pollution when developing countries open their borders to trade? Theoretical predictions are ambiguous, and empirical evidence remains limited. We study the effects of the 1991 Indian trade liberalization reform on water pollution. The reform abruptly and unexpectedly lowered import tariffs, increasing exposure to trade. Larger tariff reductions are associated with relative increases in water pollution. The estimated effects imply a 0.11 standard deviation increase in water pollution for the median district exposed to the tariff reform. 

\vspace{2ex}


\end{abstract} 

\pagebreak%
\onehalfspacing
\indent

\section{Introduction}
\label{introduction}

Global trade has expanded by a factor of 45 since the middle of the last century \parencite{wto2024}, and an increasing share of global trade involves developing countries \parencite{fouquin2016back}. How this massive expansion of trade has affected the environment in developing countries is unclear. Increased trade might have displaced polluting industries from developed countries to less-developed countries, where environmental regulations may be less stringent or less consistently enforced. This \emph{composition} effect, in the language of \textcite{grossman1991environmental}, is also known as the \emph{pollution haven hypothesis} \parencite{copeland1994north,copeland2008pollution}.  However, developed countries might export cleaner production technologies alongside polluting industries, a \emph{technique} effect that might attenuate or even reverse the composition effect. Trade also leads to growth, with its own ambiguous effects on the environment: Increased economic activity tends to create more pollution---the \emph{scale} effect---but higher incomes increase demand for environmental quality \parencite{grossman1995economic}. How trade affects the environment in developing countries is thus theoretically ambiguous, and causal empirical evidence remains sparse. 

This paper evaluates the environmental consequences of the 1991 trade liberalization reform in India. The Indian trade liberalization episode provides a unique opportunity to study the effect of openness to trade on pollution. India is by many measures among the most polluted countries in the world, with substantial costs to human health and productivity \parencite{aqli2023,worldbank2013india}. We focus on water pollution, an important determinant of agricultural production \parencite{khan2008health, mahmood2014human,sharma2006heavy,warrence2002basics}; human health \parencite{ebenstein2012consequences,garg2018not, he2016surface,do2018can}; and economic growth more generally \parencite{desbureaux2019impact}. The trade liberalization reform abruptly and unexpectedly lowered import tariff rates by, on average, almost 70\%. Districts specializing in industries with high pre-reform tariffs experienced a large shock to openness to trade, while districts exposed to lower pre-reform tariffs experienced smaller shocks.  We show that larger tariff reductions are associated with substantial relative increases in water pollution.  

To evaluate the effects of trade liberalization, we assemble a panel dataset that measures water pollution and exposure to trade between 1987 and 1997 in 117 districts, home to approximately 349 million Indian citizens at the time of the reform. We combine data on water pollution from environmental monitoring stations on rivers in India \parencite[assembled and shared by][]{greenstone2014environmental} with data on tariffs that affect exposure to trade \parencite[compiled and shared by][]{topalova2007, topalova2010factor}. We focus on eight measures of water quality---seven pollutants and dissolved oxygen---all of which are locally relevant \parencite{sargaonkar2003development} and consistently monitored throughout our study period. We estimate effects on each measure separately and on a composite water pollution index. 

We use a shift-share framework to estimate the impact of changes in tariff exposure on water pollution. Water pollution increased in districts with larger tariff reductions, compared to less-affected districts. For the median reduction of 4 percentage points in aggregate tariff exposure, water pollution increased by 0.11 standard deviations (95\% CI: 0.03, 0.20). For the individual metrics, all but one point estimate indicate increased pollution, and the remaining estimate is close to zero. We reject the null hypothesis of no relationship between tariff exposure and pollution for the water pollution index and for five of the eight individual measures of water quality at the 10 percent level or below in na\"{i}ve hypothesis tests. The estimates are robust to alternative analysis choices. The results imply that districts that are more exposed to the trade reform have substantially higher water pollution than less exposed districts. District-level census data from before and after the reform suggest that a potential mechanism may be industrialization and structural transformation away from agriculture. 

The estimated effects recover the causal impact of trade liberalization as long as changes in tariff exposure are uncorrelated with counterfactual trends in pollution. An extensive literature argues that tariff changes during the study period were not systematically related to underlying trends, finding no correlation between tariff changes and pre-reform trends in poverty and consumption, or firm- and industry-level characteristics \parencite{goldberg2010imported,topalova2007,topalova2011trade}. In our data, we likewise confirm that changes in tariff exposure are uncorrelated with pre-reform trends in water pollution. 

We find some evidence to suggest that demand for pollution abatement increased in response to trade liberalization. Districts that experienced larger tariff reductions were more likely to obtain central government funding for interventions to reduce water pollution, either in direct response to higher pollution or through more effective lobbying for national support, perhaps as a result of the economic benefits of trade liberalization. Consistent with results reported in \textcite{greenstone2014environmental}, however, these policy changes do not appear to have substantially improved water quality in the short run, although it remains possible that they improved water quality in the longer run \parencite{Lepault2024}. 

Our primary contribution is to the literature on the impact of trade on the environment \parencite{cherniwchan2017tradeb,copeland2004trade,copeland2022globalization}. A central problem in this literature is that openness to trade is not randomly assigned, leading to the potential for bias from omitted variables or reverse causality. We focus on the effects of exposure to trade on \emph{aggregate local pollution} in a developing country, an area where robust causal evidence is particularly sparse. A small number of studies link changes in trade policy in China to pollution \parencite{bombardini2020trade,chen2020wto,chen2023does} However, these studies either focus on broad national reforms with uniform effects or changes in trade policy that could be influenced by underlying industry-specific trends. In contrast, the trade liberalization reform in India affected different districts to varying degrees, and the reform was rapid, unexpected, and externally imposed. 

Our focus on \emph{aggregate} pollution complements studies that measure and decompose firm-level emissions \parencite{copeland2022globalization}. Among these studies, the most salient are \textcite{martin2011energy}, who finds that Indian firms became more fuel-efficient in response to trade liberalization but that this effect was offset by reallocation between firms, and \textcite{barrows2021foreign}, who report that growth in foreign demand for exports led firms to increase carbon emissions, a scale effect that was only partially offset by a contemporaneous technique effect. These firm-level studies typically focus on air pollutants rather than water pollutants, reflecting a broader asymmetry in the literature driven by the relative rarity of systematic and reliable data on water quality \parencite{garg2018not,excell2017thirsting}. Our focus on \emph{local} pollutants complements a related literature on the role of trade in global pollution problems, including the problem of carbon leakage \parencite[e.g.,][]{aichele2015kyoto,alsina2017effect,grubb2022carbon}. Our focus on \emph{pollution} also complements a literature on the impacts of trade on resource use \parencite[e.g.,][]{abman2020does,taylor2011buffalo}. Within this literature, the most closely related study is \textcite{sekhri2022agricultural}, who shows that groundwater levels fell rapidly in agricultural trade promotion zones in India.

Our focus on developing countries also complements a parallel literature on the effects of trade on pollution in high-income countries. Most studies in this literature find that openness to trade tends, if anything, to reduce pollution rather than increase it \parencite[see, e.g.,][]{cherniwchan2017trade,shapiro2018pollution,choi2025cleanup}, although to what extent trade can account for observed aggregate declines in pollution in high-income countries is uncertain and sensitive to measurement and accounting choices \parencite{levinson2023developed,copeland2022globalization}. In the United States, part of the observed decline in manufacturing emissions reflects firms importing dirty inputs instead of producing them in-house \parencite{cherniwchan2017trade}, displacing the associated pollution emissions abroad. This pattern of results contrasts markedly with the emerging evidence for an environmental cost to openness to trade in developing countries. 

These environmental costs must, however, be weighed against the potential economic gains \parencite[see, e.g.][]{shapiro2016trade}. Our paper thus also relates to a literature that evaluates the wider consequences of trade liberalization in developing countries and, specifically, the 1991 reform in India. In response to the reform, firms adopted new inputs \parencite{goldberg2010imported} and increased productivity \parencite{topalova2011trade}, and urban unemployment declined \parencite{hasan2012trade}. However, the economic benefits were not evenly distributed. Poverty rose disproportionately in districts more affected by the reform \parencite{topalova2007,topalova2010factor}. More-exposed children were more likely to participate in child labour and less likely to attend school \parencite{edmonds2009child, edmonds2010trade}, and they were more likely to die in infancy \parencite{anukriti2019women}. The rise in water pollution we observe could potentially help explain this increase in infant mortality \parencite[see, e.g.,][]{he2016surface,do2018can}. Our results thus extend our understanding of the broader consequences of trade liberalization in India, and in general, of the trade-offs that policymakers face when evaluating the costs and benefits of openness to trade in developing countries.  

The paper proceeds as follows. Section \ref{reform} describes the trade liberalization reform, and section \ref{data} summarizes the data. Section \ref{methodology} lays out our empirical approach, section \ref{results} presents the results, and section \ref{robustness} discusses robustness. Section \ref{conclusion} concludes. 

\section{The 1991 Indian trade liberalization reform}
\label{reform} 

\noindent India's post-independence economy was heavily centralized and regulated \parencite{cerra2002caused, khanna1995trade}. In pursuit of self-reliance, India maintained one of the most restrictive trade policies in Asia, characterized by high tariff and non-tariff barriers to trade and complex import licensing systems \parencite{banga2012twenty,cerra2002caused, topalova2010factor}. While policy began to shift gradually towards export-led growth in the latter half of the 1980s, high tariffs and import restrictions remained in place \parencite{cerra2002caused, topalova2010factor}. 

Despite these incremental reforms, India’s macroeconomic position continued to deteriorate. Fiscal imbalances grew, and current account deficits widened, driven by a confluence of adverse external and internal events: the 1990 spike in global oil prices, which was triggered by conflict in the Middle East and tripled the cost of petroleum imports; sluggish growth in important trading partner countries; and heightened political uncertainty that further lowered investor confidence \parencite{acharya2002india, cerra2002caused}.\footnote{A more detailed account of the crisis is provided in \textcite{cerra2002caused}.} Together, these pressures culminated in a balance of payments crisis in 1991.

In response to the crisis, the newly elected Indian National Congress party turned to the International Monetary Fund (IMF) and the World Bank for assistance. The international organizations provided emergency loans with the condition that India implement macroeconomic stabilization and structural reforms. In particular, they required India to remove quantitative restrictions on imports and reduce the level and dispersion of tariffs \parencite{chopra1995india, topalova2010factor}. 

The mandated reforms massively reduced import tariffs, simplified tariff and quota systems, and removed several additional import restrictions \parencite{singh2017trade}. The highest tariff rates fell from 400 percent in 1991 to 50 percent by 1995. Mean (import-weighted) tariffs fell from 87 to 27 percent over the same period, while dispersion (in standard deviations) decreased from 41 percent to 21 percent \parencite{chopra1995india}. Tariffs fell for almost all product categories, including agricultural products. The exceptions were cereals and oilseeds, which remained protected to protect food security, with imports permitted only by the state \parencite{topalova2010factor}. The share of goods subject to quantitative restrictions decreased from 87 percent in 1987 to 45 percent in 1994 \parencite{topalova2010factor}. Trade flows increased sharply after the reforms \parencite{banga2012twenty, topalova2010factor}, firms diversified their output \parencite{goldberg2010imported}, and firm productivity increased \parencite{topalova2011trade}. 

The reforms represented a drastic shift from the previous economic policy paradigm. None of the reforms were discussed during the election campaign, and the new government implemented the reforms soon after assuming power, to avoid giving opposition time to consolidate, and with little public debate, surprising even some Cabinet members \parencite{goyal1996political}. The reforms were thus sudden, unanticipated, and driven by external pressure. 

Reductions in import tariffs were initially fairly uniform across industries: Tariff changes between 1991 and 1997 were uncorrelated with pre-reform sectoral characteristics, including productivity and productivity growth \parencite{topalova2007,topalova2011trade}. \textcite{topalova2010factor} concludes that ``tariff changes between 1991 and 1997 were as unrelated to the state of the production sectors as can be reasonably hoped for in a real-world setting.'' 

The IMF-mandated liberalization policies remained in place until 1997. 
Post 1997, however, external pressure weakened, and government policy selectively targeted more productive industries during further waves of liberalization \parencite{chakraborty2014estimation, topalova2010factor}. We therefore limit our attention to the period ending in 1997.

The liberalization reforms also involved other elements, including tax reforms, revisions to restrictive industrial licensing requirements, currency devaluation, financial sector reforms, and the relaxation of constraints on foreign direct investment (FDI) \parencite{topalova2010factor,aghion2008unequal}. To the extent that currency devaluation made imports more costly, this would tend to offset any impacts of trade liberalization. That trade nonetheless increased suggests that the effects of trade liberalization dominated \parencite{banga2012twenty,topalova2010factor}. Otherwise, whether these simultaneous reforms affect our analysis depends on whether their effects were correlated in space with changes in tariffs. Following previous literature, we will test robustness of the estimated effects of trade liberalization to controls for industrial licensing and restrictions on foreign direct investment \parencite{aghion2008unequal,topalova2010factor}. 
\section{Data}
\label{data} 

\noindent We analyze the effects of trade liberalization on water pollution using a district-level panel dataset, which we assemble from multiple sources. Below, we summarize the sources from which we obtain data, how we construct variables, and how we assemble the panel dataset.

\paragraph{Tariff data} We use exposure to import tariffs to measure changes in trade policy. The data were originally collected and shared by \textcite{topalova2007, topalova2010factor}. \textcite{topalova2007} collected annual nominal ad-valorem tariff rates for approximately 5,000 product lines at the six-digit level of the Indian Trade Classification Harmonized System,   matching each product to National Industrial Classification (NIC) codes to calculate average industry-level tariffs. 

To map industry-level tariff exposure to districts, Topalova uses data on employment across industries at the three-digit NIC code level from the 1991 Indian Census and across crops within agriculture from the 1987 National Sample Survey \parencite{topalova2010factor}. Topalova then constructs district-level tariff exposure by weighting the average tariff in industry $i$ in year $t$ by the share of workers employed in industry $i$ in district $d$ at the time of the reform:

\begin{equation}
Tariff_{d,t} = \frac{\sum_{i} Worker_{d, i, t = 1991} Tariff_{i, t}}{Total~Worker_{d, t=1991}}
\label{eqn:tariff}
\end{equation}

Non-traded industries, including services, transport, construction, and cereal and oilseed cultivation, are assigned a zero tariff, both before and after the reform. In practice, the results are insensitive to the specific tariff assigned to these sectors. Since the assigned values do not change over time, they will be partialed out by district fixed effects in the panel analysis \parencite{topalova2010factor}.

\textcite{topalova2007,topalova2010factor} makes one adjustment to the raw tariff data. The raw data suggest that tariffs rebounded in 1993, but this is inconsistent with historical accounts of the reform period. We follow \textcite{topalova2007,topalova2010factor} and replace the raw 1993 data with values interpolated between 1992 and 1994.\footnote{See the raw tariff measure trend in Figure \ref{fig:trend_tariff_nonadj}. The data are identical, except for the anomalous increase in 1993. Results are robust to using the raw data.} \textcite{topalova2010factor} reports the measure of openness to trade through tariff changes for 270 districts between 1987 and 2001, out of 339 districts in total defined by their 1966 boundaries.

\paragraph{Pollution data} Water quality is measured in samples collected at monitoring stations in freshwater bodies throughout India, including rivers, lakes, canals, and aquifers. The monitoring network is managed by the Indian Central Pollution Control Board (CPCB) and its state counterparts. The water quality data were assembled and shared by \textcite{greenstone2014environmental}. We focus on data from monitoring stations on or by rivers, following \textcite{greenstone2014environmental} who argue that such stations are more consistently monitored. Monitoring stations mostly report data on a monthly or quarterly basis (Figure \ref{fig:pollution_count}). The first monitoring stations opened in 1986. In total, data are recorded at 468 monitoring stations along 162 rivers at least once between 1986 and 2005.

The full dataset contains data on many different water quality metrics, but not all are consistently recorded across space and time. We focus on metrics that environmental engineers identify as critical elements of a comprehensive India-specific water quality index \parencite{sargaonkar2003development}.\footnote{Further details are in Appendix \ref{app:pollutants}. \textcite{sargaonkar2003development} include four water quality metrics that we omit because of data availability: color is not included in the pollution dataset, and arsenic, nitrate, and fluoride have few non-missing observations ($>$85\% missing at the station level). We also omit pH because almost all observations (96\%) fell in the acceptable range. See: https://cpcb.nic.in/water-quality-criteria. Changes to pH are also difficult to interpret because both low and high levels are damaging.} Specifically, we focus on seven pollutants: \emph{Biochemical oxygen demand} is a measure of organic pollution. \emph{Turbidity} and \emph{total dissolved solids} reflect the quantity of particulate matter dissolved or suspended in water. \emph{Total coliform bacteria} measures the presence of a broad group of bacteria, some of which thrive in the intestines of warm-blooded animals. \emph{Sulfates} have a range of anthropogenic sources, largely industrial, while anthropogenic sources of \emph{chloride} include agriculture, energy production, and wastewater. \emph{Hardness} measures the content of dissolved minerals. Although we include hardness in our analysis, following \textcite{sargaonkar2003development}, we note that it is primarily driven by local geology rather than anthropogenic pollution \parencite{carr2008water}. We also include one measure of general water quality, \emph{dissolved oxygen} (DO). To facilitate comparison of effect sizes, we standardize all water quality metrics relative to their pre-reform (pre-1991) mean and standard deviation, and we reverse the scale for DO so that, as for the pollutants, higher values are associated with lower water quality. 

We also summarize information about the individual water quality metrics in a water pollution index that captures overall changes in water pollution, using inverse covariance weighting \parencite{anderson2008multiple}. The inverse-covariance-weighted index downweights highly correlated variables, maximizing the information contained in the index. We calculate the index across all non-missing water quality metrics for each district-year observation.\footnote{Following \citeauthor{anderson2008multiple}'s (\citeyear{anderson2008multiple}) formulation, we adapt code provided by \textcite{samii} to deal with cases where not all water quality metrics are reported for a given district and year. See Appendix \ref{app:anderson_index} for details. Table \ref{table:corr} shows the correlations between the water quality metrics. } We report the effects of trade liberalization both for the pollution index and the individual metrics.  

\paragraph{Panel} We first collapse the pollution data to district-year average values.\footnote{We also confirm results are robust to using monitoring-station level data.} We then match the district-level pollution data to the tariff data using district names. District names differ in the two datasets in some cases, primarily because of spelling or naming conventions. We use information on district name and boundary changes from \textcite{kumar2017creating}. We limit our sample to 137 districts with pre-reform pollution data (i.e., from 1990 or earlier). After excluding one district where pre-reform tariffs were 107\% larger than in any other sample district,\footnote{Our main results remain consistent in sign and statistically significant if we include this district, but the district has a \citeauthor{Cook1977}'s (\citeyear{Cook1977}) distance of 23, suggesting very high leverage.} we successfully match 117 districts with pre-reform pollution data to tariff exposure data (see Figure \ref{fig:water_map}), 

\begin{sidewaysfigure}[p]

    \includegraphics[width=
    \textwidth]{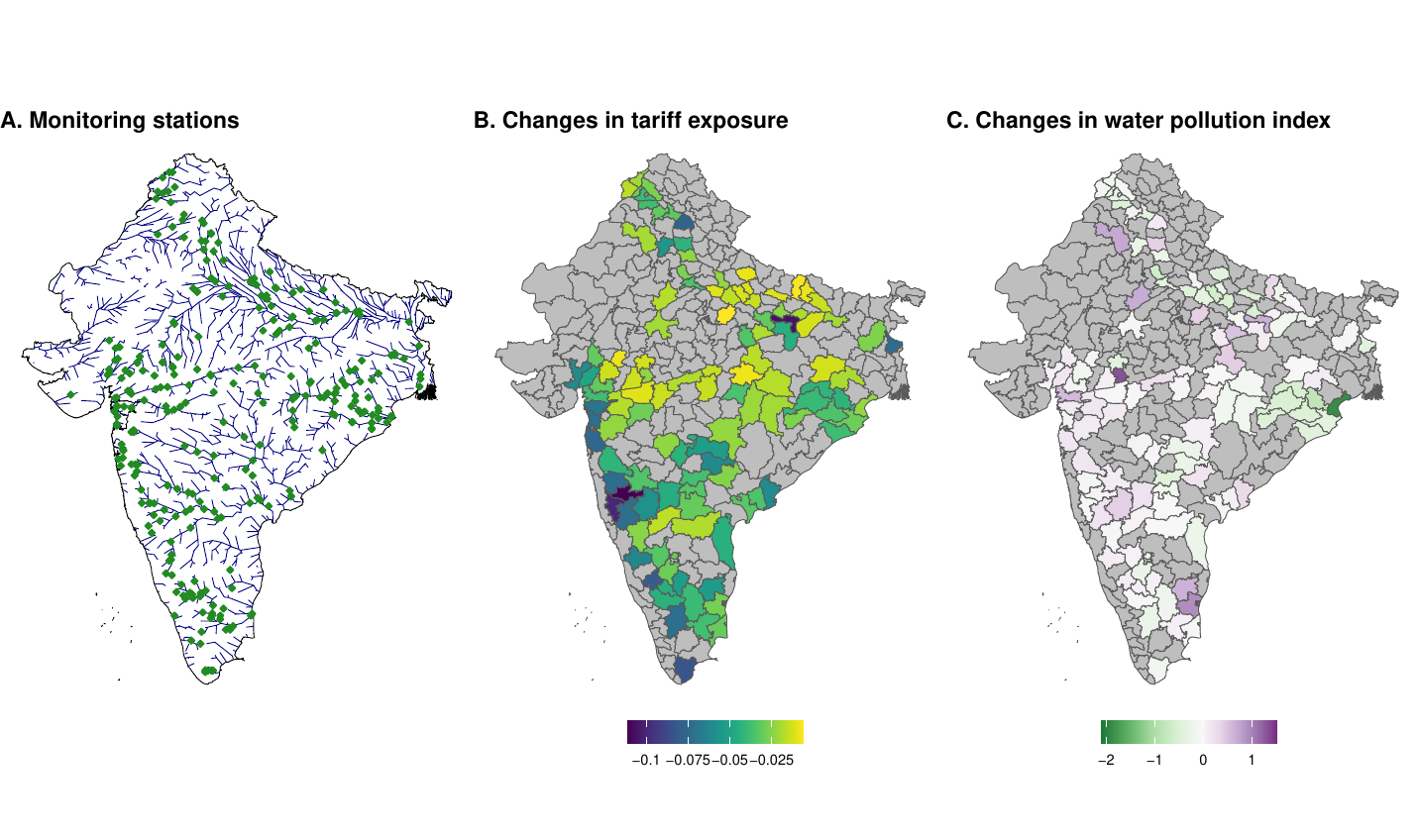}
    \caption{\raggedright Panel A shows larger rivers (blue) along with monitoring stations (green); panel B shows district-level changes in tariff exposure, where more negative values imply greater increases in exposure to trade; and panel C shows district-level changes in the water pollution index, where positive values (purple) indicate increases in water pollution and negative values (green) decreases. Changes in variables are measured as the difference between average values before and after 1991. Districts that are not in the sample are shaded grey. District boundaries are drawn as defined in 1966. We exclude the Andaman and Nicobar Islands, Jammu and Kashmir, and the northeastern states, since data are missing for all districts in these regions. Data on district boundaries are from \textcite{MLInfomap2013}, and \textcite{IPUMS_India_1987_GIS}. River network data is from \textcite{lehner2013global}.}
    \label{fig:map}
    
\end{sidewaysfigure}

The final panel thus has data on 117 districts between 1987 to 1997, corresponding to pollution data from 260 monitoring stations. Truncating the sample in 1997 limits our analysis to the period during which tariff changes were uncorrelated with industry productivity \parencite{topalova2010factor,topalova2011trade}. Panel A in Figure \ref{fig:map} maps the monitoring stations included in the final sample. Panels B and C show changes in tariff exposure and pollution over the study period, respectively. While not comprehensive, the dataset covers large swathes of India, in total 34\% of its land surface area. In the 1991 Census, the recorded population of the sample districts was approximately 349 million, about 41\% of the total population. Census data from the time of the reform suggests that the sample districts on average do not systematically differ from the districts that are excluded from our sample because of missing data.\footnote{See Table \ref{table:balance_early}. Some differences between the sample and the excluded districts are statistically significant, but they are neither large in magnitude nor systematic. Sample districts are slightly more urbanized but have a slightly larger share employed in agriculture. The apparent statistical significance of these differences may simply reflect spatial correlation in census data outcomes, and tariff and pollution data availability.}

\begin{figure}[ht!]
    \centering
    \includegraphics[width=0.85\textwidth]{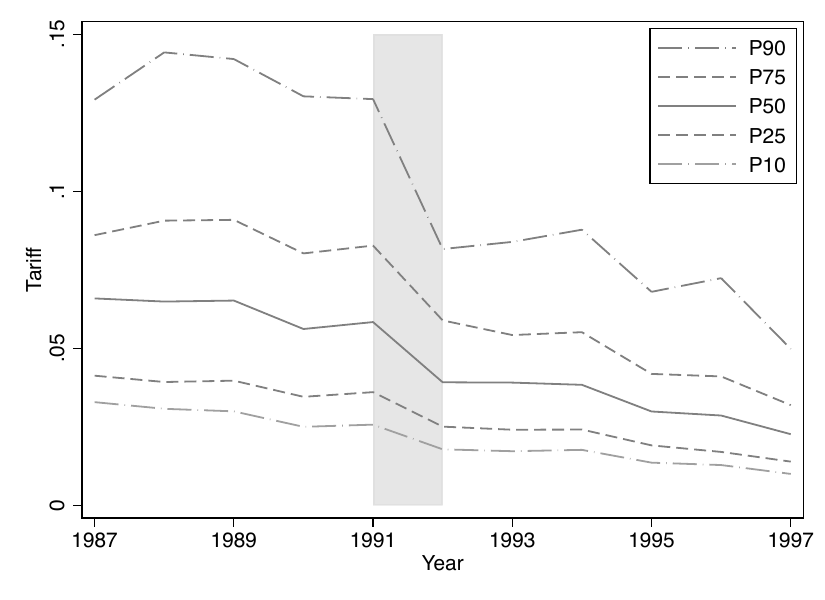}
    \caption{\raggedright Distribution and trend of district-level tariff exposure.  Lines show the listed percentile of the data. Shaded area indicates the reform period. Data for 1993 are imputed following \textcite{topalova2007, topalova2010factor}.}
    \label{fig:trend_tariff_percentiles}
\end{figure}

Trends in tariff exposure are plotted in Figure \ref{fig:trend_tariff_percentiles}. Tariff exposure declines for all districts in our sample, but the magnitude of the decline varies substantially, ranging between 2 and 10 percentage points (p.p.), with tariffs declining by 4 p.p. for the median district. The decline is steepest between 1991 and 1992 immediately after the reform.\footnote{Figures \ref{fig:hist_tariff_y} and \ref{fig:hist_tariff_Dy} plot the distribution of the tariff exposure variable across districts and years, and the distribution of changes in tariff exposure across the study period, respectively.}

\begin{figure}[t!]
   \centering

   \begin{subfigure}{0.72\textwidth}
     \includegraphics[width=\linewidth]{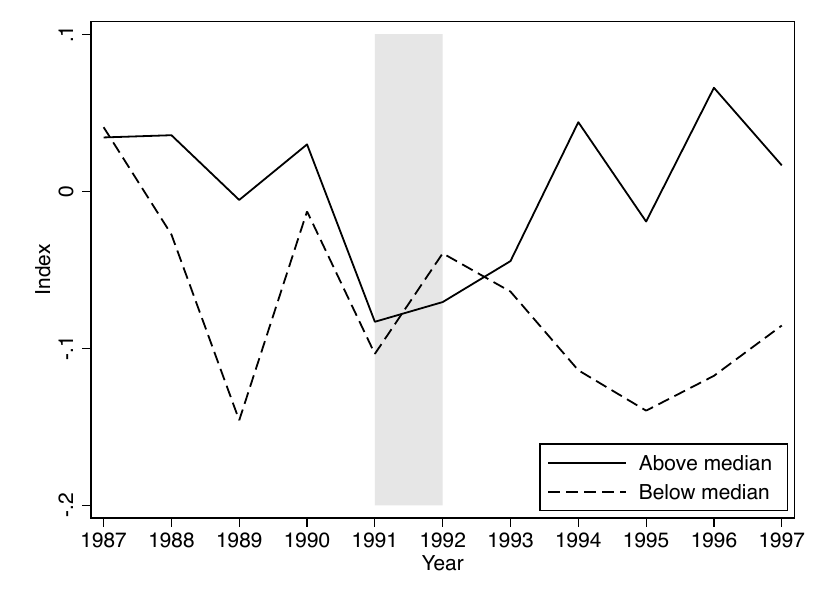}
     \caption{Trends in pollution index by changes in tariff exposure}
     \label{fig:trend_index}
   \end{subfigure}
   \begin{subfigure}{0.72\textwidth}
     \includegraphics[width=\linewidth]{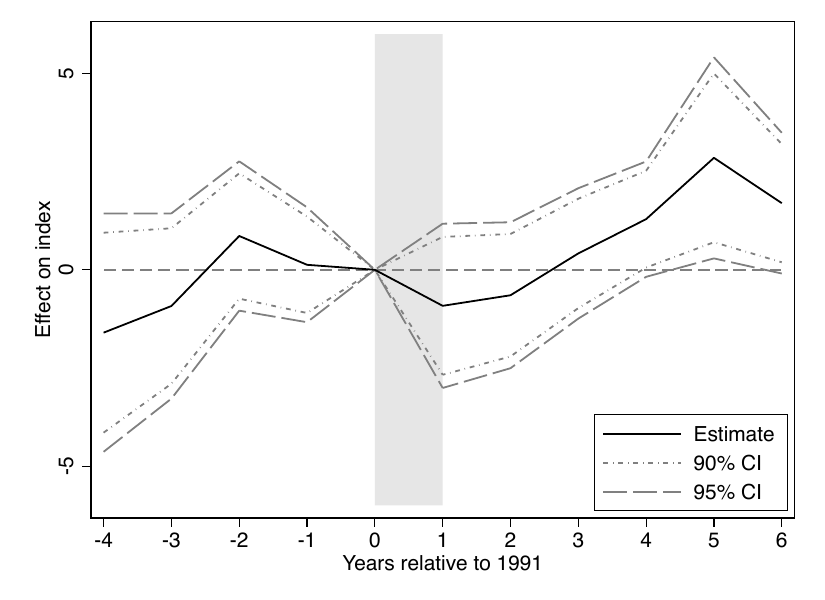}
     \caption{Correlation between pre-form tariff exposure and trends in pollution index}
     \label{fig:es_index}
   \end{subfigure}

   \caption{\raggedright a) Trends in water quality, averaged across districts with above and below median changes in tariff exposure (in absolute magnitude). b) Point estimates from a regression of the pollution index on pre-reform average tariffs, interacted with event year indicator variables, omitting 1991, and controlling for district and year fixed effects. Shaded area indicates the reform period.} 
  
\end{figure}

Figure \ref{fig:trend_index} plots trends in water quality over the study period, for districts with above and below median tariff exposure changes. Districts that later experience smaller and larger changes in tariff exposure have quite similar pollution levels and trends before the reform.\footnote{Table \ref{table:lvl_results} confirms that, in the cross section, future changes in tariffs are significantly correlated with pre-reform values of only one of the eight water quality metrics (DO). Figure \ref{fig:trend_all_pollutants} shows trends for districts with above and below median tariff exposure changes for all individual water quality metrics, and Figure \ref{fig:hist_water} shows the distribution of the values for each water quality metric.} Figure \ref{fig:es_index} shows how the correlation between pre-reform tariffs and pollution evolves over time.\footnote{Specifically, Figure \ref{fig:es_index} shows coefficients from estimating the following equation, clustering by district: \begin{equation} 
Y_{d,t} = \alpha + \sum_{s=-4}^{-1} \beta_s Tariff_{d,pre} \mathbf{1}_{t-1991 = s} + \sum_{s=1}^{6} \beta_s Tariff_{d,pre} \mathbf{1}_{t-1991 = s} + \gamma_d + \delta_t + \varepsilon_{d,t}
\label{eqn:es_model}
\end{equation}}  Before trade liberalization, pre-reform tariffs were weakly and statistically insignificantly correlated with the water pollution index, conditional on time and district fixed effects. Following the reform, trends in pollution start to diverge. The difference in trends associated with pre-reform tariffs is statistically significant towards the end of the sample period, when tariff reductions are largest. Each individual coefficient shown in Figure \ref{fig:es_index} is estimated imprecisely, however. Although the exercise shows no obvious or systematic divergence in pre-reform trends, we cannot rule out the possibility that the data-generating process that produces the pre-reform coefficients could also produce the apparent post-reform effects \parencite{rambachan2023more}. This reflects an intrinsic limitation of the setting, in which pollution data are only available for four pre-reform periods. Our main analyses will pool effects over time and account for the exact timing and magnitude of tariff reductions, giving us greater statistical power both to detect effects and to conduct falsification exercises. 

Although \textcite{greenstone2014environmental} also provide air pollution data, we focus on water pollution because only 29 districts had both non-missing pre-reform air pollution data and matched tariff data.

\paragraph{Auxiliary data} We use additional data to explore potential mechanisms and evaluate robustness. To explore potential mechanisms, we use data from the National Sample Survey and Primary Census Abstract that has been cleaned and harmonized by \textcite{liu2023climate}, and data from the Economic Census \parencite{ecindia}, made available by \textcite{asher2021development}. We also use additional data on environmental policies from \textcite{greenstone2014environmental} to explore whether districts enact policies designed to reduce water pollution in the wake of trade liberalization. We further evaluate robustness to controlling for other policies implemented during the reform period using data on non-tariff barriers, foreign direct investment, and licensing in a subsample of districts and years, provided by \textcite{topalova2010factor}. When we use these auxiliary data sources, the sample size sometimes varies as a result of spatial and temporal data availability in the auxiliary sources and imperfect matching across sources.

\section{Empirical strategy}
\label{methodology} 

\noindent This section describes how we estimate the effects of trade liberalization on water pollution. Our approach draws on \textcite{topalova2007, topalova2010factor} and subsequent literature. Tariffs were lowered for different industries to different extents at different times, and industrial composition varied across districts. The reform thus affected districts with varying intensity. 

Comparing districts that experienced greater tariff reductions to those that experienced lower tariff reductions, we estimate the following equation using OLS:

\begin{equation} 
Y_{d,t} = \alpha + \beta  Tariff_{d,t} + \gamma_d + \delta_t + \varepsilon_{d,t}
\label{eqn:main_model}
\end{equation} 

\noindent where: $Y_{d,t}$ is a measure of water pollution in district $d$ and year $t$;  $Tariff_{d,t}$ is district-level tariff exposure, as in Equation \ref{eqn:tariff}; and $\beta$ is the coefficient of interest, which captures the estimated effect of tariff changes on water pollution. The district fixed effects $\gamma_d$ absorb all time-invariant differences across districts. The year fixed effects $\delta_t$ absorb national trends. 

The parameter $\beta$ has a causal interpretation as long as changes in Indian trade policy are uncorrelated with trends in water pollution through any other channel but the causal effect. Because district fixed effects absorb pre-reform differences that are correlated with initial tariff exposure, this is equivalent to assuming that the changes in tariff exposure are exogenous to counterfactual changes in pollution \parencite[see, e.g.,][]{borusyak2022quasi}. 

The validity of this identifying assumption is supported by the sudden, unanticipated, and externally mandated nature of the trade liberalization reform. Using data from \textcite{topalova2010factor}, we confirm that post-reform changes in our primary measure of tariffs are uncorrelated with pre-reform trends in poverty and consumption (Table \ref{table:topalova_fals}). Other  studies further confirm that changes in tariffs during the study period were uncorrelated with firm and industry characteristics \parencite{goldberg2010imported,topalova2011trade}.   

As an additional test of the validity of the identifying assumption, we evaluate whether pre-reform pollution trends are correlated with future tariff changes in our data. We re-estimate Equation \ref{eqn:main_model}, replacing $Tariff_{d,t}$ with $Tariff_{d,t+4}$. In other words, we test whether future changes in the tariff measure predict past changes in water pollution. Since any such relationship could not be causal, rejecting the null hypothesis of no relationship would raise concerns about the validity of our identifying assumption. 

Some studies of the trade liberalization episode use an instrument for tariff exposure that is calculated only across traded sectors, either as a robustness test or as their main specification. The motivation for using this instrument is the possibility that trends could differ across districts with different shares of non-traded industries. However, it is not clear that this alternative approach is preferable. The instrument does not always pass modern weak instrument tests, and using the instrument does not typically change the sign of the estimated effect but decreases precision.\footnote{An exception to this are \citeauthor{anukriti2019women}'s (\citeyear{anukriti2019women}) results on fertility, although their results on mortality are insensitive to instrumenting for tariff reform.} In addition, in \citeauthor{topalova2010factor}'s (\citeyear{topalova2010factor}) data, although the falsification tests that simply use tariff exposure are reassuring, falsification tests \emph{fail} in some samples when using the non-traded-sector instrument (Table \ref{table:topalova_fals_IV}).   

In our sample of districts with water pollution data, an instrument based on tariff rates in traded sectors passes weak instrument tests but explains only a small share (partial $R^2$ = 0.16)  of the observed variance in tariff exposure (Figure \ref{fig:fs} and Table \ref{table:iv_results}). If we use traded tariff exposure to instrument for tariff exposure in our regression framework, we obtain much less precisely-estimated coefficients with confidence intervals that do not exclude zero (Figure \ref{fig:ols_iv}). We cannot reject the null that the OLS and IV estimates are equal for the index and for 7 out of 8 of the individual pollution metrics.\footnote{The exception is chloride, where we reject the null of equality of the OLS and IV coefficients with $p$=0.07 (unadjusted for multiple hypothesis testing).} Since the IV approach sacrifices power and precision without offering obvious advantages in terms of bias reduction, we focus on the OLS specifications throughout the paper. 

We cluster standard errors at the district level, allowing for arbitrary serial correlation in the residual term $\varepsilon_{d,t}$. Spatial correlation in tariff reductions and trends in pollution appears fairly low in our sample---see panels A and B in Figure \ref{fig:map}---and results are very similar if we additionally account for spatial correlation in the error term.  

We measure effects on aggregate outcomes at the district level. Our results thus capture the aggregate effects of trade liberalization within a district, or more specifically, the net effect of any scale, composition, or technique effects.

We primarily focus on tariff exposure throughout the analysis. Tariffs can be measured consistently over time \parencite{goldberg2004trade} and are available at a disaggregated level for our full study period, whereas comparable data on non-tariff barriers to trade (NTBs) are not \parencite{topalova2007}. Since tariffs and NTBs are correlated \parencite{goldberg2004trade,topalova2007}, our tariff measure likely captures the broader effects of trade liberalization rather than the specific effects of tariff reductions alone. In a subset of our sample for which data on NTBs are available  from \textcite{topalova2010factor}, controlling for NTBs does not change the estimated effect of tariffs, but this may simply reflect the relative difficulty of measuring NTBs consistently rather than the specific importance of tariffs in driving the effects. 

Our approach evaluates whether changes in water pollution differed between districts that experienced large tariff reductions and those where tariffs remained relatively stable. This design does not capture any effects of trade liberalization---or of any other reforms---that affected districts uniformly. However, districts with relatively small changes in tariff exposure did not experience obvious changes in water pollution trends during the post-reform period (Figure \ref{fig:trend_index}). 

In principle, relative changes in pollution could be driven by the displacement of pollution across districts within India. Our analysis does not allow us to test whether this is the case. However, tariffs were constant within industries across districts. The reform thus did not incentivize the movement of industrial activity across districts. \textcite{topalova2010factor}  finds no evidence that trade liberalization induced migration across districts. Using census data assembled by \textcite{liu2023climate}, we also find no evidence that population grew differently in districts more and less exposed to the reform. It thus seems unlikely that the displacement of pollution across space could explain our results. 

\section{The impact of trade liberalization on water pollution}\label{results} 

\begin{figure}[ht!]
    \centering
    \includegraphics[width=1\textwidth]{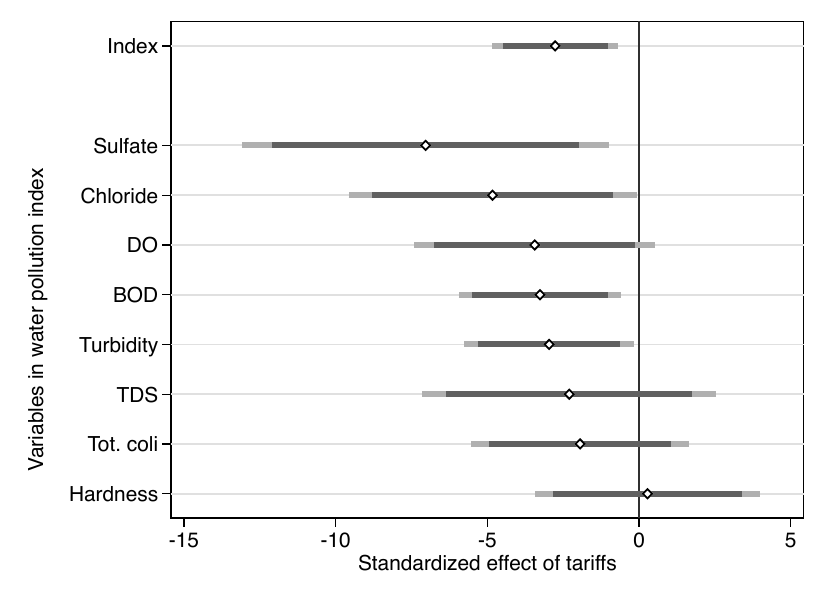}
    \caption{\raggedright Results from the main regression specification, where the estimates show the standardized effect of a contemporaneous change in the tariff measure. More extreme values suggest larger effects of trade on pollution.  For DO, the scale is reversed, since DO negatively correlates with pollution levels. District and year fixed effects included in all regressions. Standard errors are clustered by district. Dark grey confidence intervals at the 90\% level; lighter grey confidence intervals at the 95\% level.}
    \label{fig:main_results}
\end{figure}

The results show that water pollution increases in response to trade liberalization. Table \ref{table:main_results} and Figure \ref{fig:main_results} summarize the results. Each point estimate is obtained by separately estimating Equation \ref{eqn:main_model} for the water pollution index or the individual water quality metrics. Since tariffs are negatively correlated with trade, a negative relationship between tariffs and pollution implies that trade increases pollution. We reverse the scale for DO for consistency with the other outcomes. For the water pollution index, as well as for four (sulfate, chloride, BOD, turbidity) of the seven individual pollutants, we reject the null hypothesis of no relationship between trade regulation and water pollution at the 5 percent level. We also reject the null hypothesis of no relationship for DO at the 10 percent level.  

All outcome variables are standardized, so each point estimate reflects the change in pollution in standard deviations were tariff exposure to fall by 100 percentage points. In the median district, tariff exposure fell by 4 percentage points in the post-reform period. Scaling the point estimates by this change suggests that the water pollution index increased for the median district by 0.11 standard deviations (95\% CI: 0.03, 0.20). The equivalent effects on the individual metrics range from an increase of 0.29 s.d. (95\% CI: 0.04, 0.54) for the most affected pollutant (sulfate) to a decrease of 0.01 s.d. (95\% CI: -0.16, 0.14) for the least affected pollutant (hardness). That hardness is relatively unaffected may be reassuring, since among the pollutants we study it is the least connected to human activity. 

The number of observations varies across pollutants because of missing data. Among district-years with any pollution data, the share of missing observations varies from 0.5\% (DO) to 18.8\% (TDS). The pollutants interpreted by previous research as useful general proxies for water quality \parencite{greenstone2014environmental, do2018can, keiser2019consequences}, specifically BOD and DO, have fewer missing observations, narrower confidence intervals, and point estimates that lie in the middle of the effect size distribution. 

We also estimate a stacked regression at the water quality metric-district-year level, allowing fixed effects and coefficients to vary by water quality metric. The stacked regression replicates the point estimates from the individual regressions and allows us to compare effect sizes for different water quality metrics. Despite the variance in point estimates, we cannot reject the null hypothesis that the effects are equal for all metrics ($p$ = 0.188). We reject the null that coefficients jointly equal zero for all water quality metrics ($p$ = 0.026).

\begin{figure}[t!]
    \centering
    \includegraphics[width=1\textwidth]{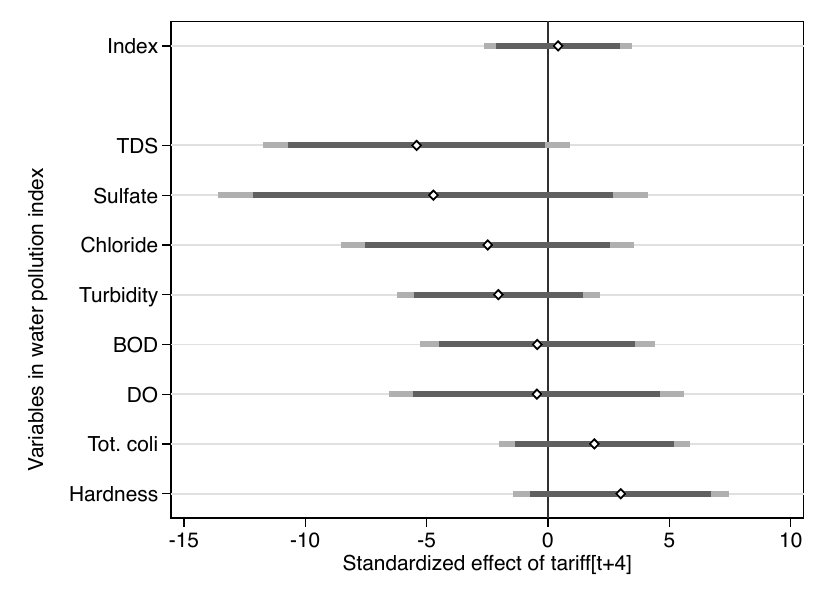}
    \caption{\raggedright Results from the falsification exercise, where each water quality metric and the water pollution index are regressed on a four-year lead of the tariff measure. More extreme values suggest larger effects of trade on pollution. For DO, the scale is reversed, since DO negatively correlates with pollution levels. District and year fixed effects are included in all regressions. Standard errors are clustered at the district level. Sample includes data for years between 1987 and 1993. Dark grey confidence intervals at the 90\% level; lighter grey confidence intervals at the 95\% level.}
    \label{fig:falsification_all}
\end{figure}

The results of the falsification test described in Section \ref{methodology} are shown in Figure \ref{fig:falsification_all}.\footnote{See also Table \ref{table:fals_table_all}.} The test evaluates whether future tariff changes are correlated with pre-reform pollution trends. The sample size is smaller for the falsification test, and the point estimates are noisier and less precisely estimated, but they are on average closer to zero than the estimated treatment effects. None of the 95\% confidence intervals exclude zero for any of the water quality metrics or the pollution index; the 90\% confidence interval excludes zero for one individual pollutant (TDS, $p$=0.092 with no adjustment for multiple hypothesis testing). The relatively precise point estimate for the water pollution index has the opposite sign from the main treatment effect in the falsification test and is 85\% smaller in absolute magnitude. In a stacked regression, we reject the null that the coefficient from the falsification exercise is equal to the estimated effect of tariffs on the water pollution index ($p$=0.031). These results provide some reassurance about the validity of the identifying assumption.  

\begin{sidewaystable}[p]
    \centering
    \begin{threeparttable}
        \caption{The effect of trade liberalization on water pollution}
        \input{tables/table1}
        \label{table:main_results}
        \begin{tablenotes}
            \small
            \item Notes: Standard errors (in parentheses) are clustered at the district level. The index and all measures of water pollution are normalized to the pre-reform period. Year and district fixed effects are included in all regressions. \mbox{* $p$ $<$ 0.10, ** $p$ $<$ 0.05.}
        \end{tablenotes}
    \end{threeparttable}
\end{sidewaystable}

In summary, reduced tariffs---and consequently increased exposure to trade---increased water pollution. 


\paragraph{Potential mechanisms} Trade potentially affects pollution through a broad range of channels, including scale, composition, and technique effects \parencite{grossman1991environmental}. Were firm-level water pollution data available, one could apply approaches from the decomposition literature to disentangle these effects \parencite[see, e.g.,][]{shapiro2018pollution}. However, firm-level water pollution data are extremely scarce and absent for our study period in India. 

To shed what light we can on mechanisms, we turn to census data \parencite{asher2021development,liu2023climate}. With the important caveat that we only observe districts twice during our study period in these data, once pre- and once post-reform, which limits statistical power,\footnote{Studies that successfully detect effects of the trade liberalization episode often work with either higher frequency data \parencite{anukriti2019women} or more granular data that allows them to exploit differences in exposure across firms or industries \parencite{goldberg2010imported,topalova2011trade}.} we find that employment shifts strongly away from the agricultural sector towards the non-agricultural sector. While population growth appears largely stable, urbanization accelerates. Point estimates suggest that trade increases total employment and employment in all non-agricultural sectors, and the number of privately-owned firms, although none of these relationships are precisely estimated (Tables \ref{table:pca} to \ref{table:ec_firm_m}).

Because we focus on general measures of water quality, most of the pollutants we study have multiple sources, often including agricultural, domestic, and industrial sources \parencite{chapman1996water,chapman2013surface}, meaning that heterogeneous effects across pollutants are not strongly informative about different potential mechanisms. However, the strongest effect we observe is for sulfate, which is also the pollutant that is most commonly linked to industrial activity. Overall, the results are consistent with an increase in pollution driven by industrialization and structural transformation away from agriculture. 

\paragraph{Demand for environmental quality} Trade may increase demand for environmental quality, either in response to pollution or as a result of higher income \parencite{grossman1995economic}. We evaluate whether trade liberalization leads districts to enact policies designed to reduce pollution, using data from \textcite{greenstone2014environmental}. The policies in question are pollution abatement interventions, implemented initially in the Ganges River Basin before being scaled up across India through the National River Conservation Plan (NRCP). Under the NRCP, the central government provides funds and technical support for interventions to improve water quality in polluted rivers, such as sewage treatment works. State governments propose interventions and sites; whether their proposals are funded depends on factors including the degree of pollution and the availability of funds \parencite{india_nrcp_lok_sabha_2018}. We estimate Equation \ref{eqn:main_model} for an additional outcome variable $Policy_{d,t}$, equal to 1 if a pollution abatement intervention was enacted in any city within a district $d$ in or before year $t$. We find that decreases in tariffs increase the share of districts with abatement policies (column 1 of Table \ref{table:policy_results}).  At the median reduction in tariff exposure, the result is equivalent to an 11\% (95\% CI: 0.00, 0.21) increase in the share of districts with any policy in place.

The interventions did not appear to effectively reduce pollution in the short run, however, consistent with results in \textcite{greenstone2014environmental}, who also find that these interventions had no discernible effects on pollution in a larger sample over a longer period. We augment Equation \ref{eqn:main_model} to interact tariff exposure either with $Policy_{d,t}$ or with a dummy that equals one if district $d$ ever implemented a policy to reduce water pollution and zero otherwise. With the caveat that policy reforms are not exogenous, we find no evidence to suggest that policies mediated the effect of tariffs, nor that pollution evolved differently in districts that enacted policies. Column 2 of Table \ref{table:policy_results} restates the main results for comparison. We then show that the pollution index is uncorrelated with $Policy_{d,t}$, conditional on year and district fixed effects, and that the effect of the tariff reform does not depend on $Policy_{d,t}$ (column 3). We also cannot reject the null hypothesis that the response to tariffs is the same in districts that ever enacted policies and those that did not (column 4).

\begin{table}[t!]
    \centering
    \begin{threeparttable}
        \caption{Trade liberalization and environmental policy}
        \input{tables/table2}
        \label{table:policy_results}
        \begin{tablenotes}
            \small
            \item Notes: Standard errors (in parentheses) are clustered by district. The index is normalized to the pre-reform period. Year and district fixed effects are included in all regressions. \mbox{* $p$ $<$ 0.10, ** $p$ $<$ 0.05}.
        \end{tablenotes}
    \end{threeparttable}
\end{table}

\section{Robustness and sensitivity}
\label{robustness} 

\begin{figure}[p!]
    \thisfloatpagestyle{empty}
    \centering
    \includegraphics[width=1\textwidth]{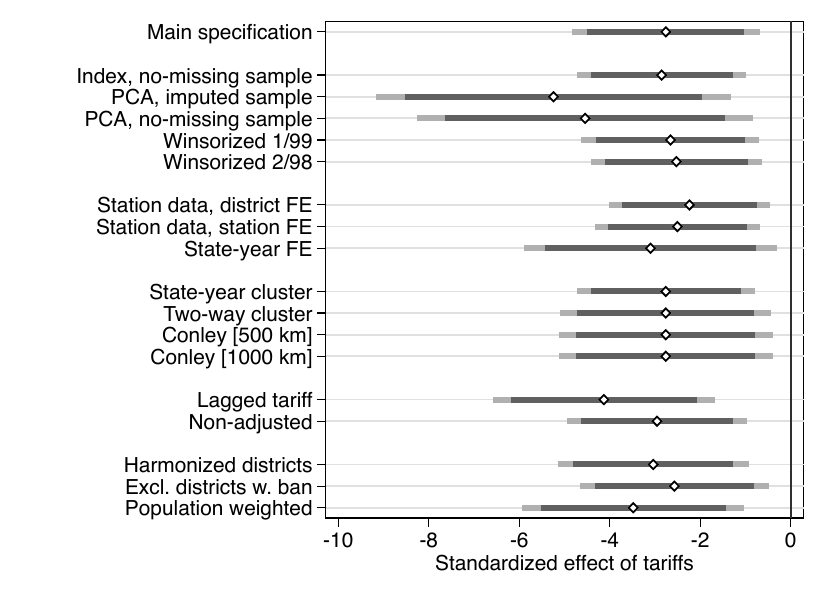}
    \caption{\raggedright \textbf{Robustness} Main specification: inverse covariance-weighted index, contemporaneous effects, district and year fixed effects, standard errors clustered by district. Robustness tests vary analysis choices, as follows: Index, no-missing sample: only district-year observations with no missing data for any metrics. PCA, imputed sample: PCA index, with missing variables imputed. PCA, no-missing sample: PCA index, only district-year observations with no missing data for any metrics. Winsorized: extreme 2 or 4 percentiles of pollution data winsorized. Station data, district FE: main specification using monitor station-level data. Station data, station FE: monitoring-station fixed effects. State-year FE: global trends modeled with state-year fixed effects. State-year cluster: standard errors clustered by state-year. Two-way cluster: standard errors clustered two-way by district and year. Conley: Conley standard errors in addition to clustering by district, over 500 and 1000 km respectively. Lagged: effects of tariffs lagged one year. Non-adjusted: tariff data without adjusting value in 1993. Harmonized districts: unit of analysis is a harmonized district  \parencite{liu2023climate}, with area-weighted tariff and pollution variables. Excl. districts w. ban: drops 3 districts affected by a local ban on leather tanneries. Population weighted: using district population in 1991 as regression weights. Dark grey confidence intervals at the 90\% level; lighter grey confidence intervals at the 95\% level. N = 1,146 observations (2,481 at the station level) in 117 districts, except in non-missing sample (N = 740 in 99 districts), with lagged tariffs (N = 1,029 in 117 districts), harmonized districts (N = 950 in 96 districts), and excluding districts with the tannery ban (N = 1,117 in 114 districts).}
    \label{fig:robustness_index}
\end{figure}

The results are robust to a series of alternative analysis choices, including different ways to measure water pollution, specifications, approaches to inference, ways to measure tariff exposure, and approaches to sampling and weighting. In the interests of brevity, we summarize the results of robustness tests for the water pollution index in Figure  \ref{fig:robustness_index}.  Detailed corresponding results for the individual water quality metrics are in Figures \ref{fig:robustness_bod} to \ref{fig:robustness_turb}.
  

\paragraph*{Measuring water pollution} Our main analyses use inverse covariance weighting \parencite{anderson2008multiple} to construct the water pollution index.  An alternative approach to summarizing information about multiple outcomes is principal components analysis (PCA). If we use a PCA index to summarize effects on water pollution, the estimates are larger and less precisely estimated.\footnote{Figure \ref{fig:index_distribution_comp} compares the distribution of the two indexes.} Unlike the inverse covariance weighting approach, PCA cannot be used with missing observations, so we impute the values for missing water quality metrics using the values of the non-missing metrics.  Results for both indexes are also very similar if we limit the sample to district-years with no missing observations for any individual metrics.

The distributions of some of the individual water quality metrics are skewed (Figure \ref{fig:hist_water}). However, the results remain very similar if we winsorize the outcome variables either at the 1st and 99th percentiles, or at the 2nd and 98th percentiles.

To evaluate sensitivity to the choice of water quality metrics to include in the index, we create alternative indexes that sequentially exclude each water quality metric. The estimated effects remain very consistent in both magnitude and significance when we exclude any individual metric (Figure \ref{fig:coefplot_loo}). 

\paragraph*{Alternative specifications} Our main specification uses district-level data obtained by averaging measures of pollution across monitoring stations. The network of water pollution monitoring stations expanded during our study period (Figure \ref{fig:yop}). Most of the expansion took place before the trade liberalization reform, making it unlikely that heterogeneous changes in monitoring could affect our results. Using monitoring station-level data instead of district-level data yields similar results for the pollution index, either with district fixed effects, as in the main specification, or with monitoring-station fixed effects. The point estimates are also extremely stable if we progressively limit our sample to monitoring stations or districts with data from earlier in our sample period (Figure \ref{fig:missingness}).

One might further be concerned that the results could be driven by differences in regional trends within India. However, the estimated effect on the water pollution index is larger, albeit less precisely estimated, if we allow trends to vary across regions by including state-year fixed effects in the main specification. One caveat on sensitivity is that the results are not robust to including district-year trends. However, district-year trends absorb a substantial share of the variance in tariff exposure that we use to estimate effects. The results remain very stable if we allow trends to vary flexibly with respect to any of a wide range of pre-reform characteristics measured in census data (Figures \ref{fig:initial_nss} to \ref{fig:init_ec}). 

The reforms we study included other changes to trade policy, such as changes to non-tariff barriers to trade (NTBs), as well as other simultaneous structural reforms, including the removal of restrictions on foreign direct investment and industrial delicensing \parencite{aghion2008unequal,topalova2010factor}. \textcite{topalova2010factor} provides data on these policy changes for a subset of districts in 1987 and 1997. Restricting our sample to these districts and years, we obtain similar results to our main specifications. Adding controls for NTBs or other reforms tends, if anything, to increase the estimated effect of tariffs (Table \ref{table:ntb}). Given that NTBs are harder to measure than tariffs, as discussed in Section \ref{methodology}, one may still wish to interpret our results primarily as a generalized effect of trade rather than a specific effect of reduced tariffs. However, the stability of the effects of trade liberalization when controlling for other simultaneous reforms suggests these other reforms are unlikely to explain the observed effects. 

\paragraph{Modelling standard errors} Our main specifications cluster by district, accounting for arbitrary potential serial correlation of outcomes within districts.  The estimated standard errors are very similar if we alternatively cluster by state-year \parencite[as in][]{topalova2010factor}, two-way cluster by both district and year, or  model \textcite{conley1999gmm} standard errors in addition to clustering by district. For the individual water quality metrics, a few marginally change significance using these alternative approaches to inference, but the effect on the index is consistently statistically significant with any of them.   
\paragraph{Measuring tariff exposure} Our main specification estimates the effect of contemporaneous tariff exposure on pollution. This model could be mis-specified if effects take time to materialize. Lagging the tariff exposure measure by one year so that we estimate the effects of tariff exposure in year $t-1$ on pollution in year $t$ indeed yields fractionally stronger results.  

Our main estimates follow \textcite{topalova2007, topalova2010factor} and use an interpolated measure of tariff in 1993. In practice, using the raw tariff data yields very similar results. 

\paragraph{Harmonized district boundaries} The units of observation in our main analysis are districts defined by their 1966 boundaries, as in \textcite{topalova2007,topalova2010factor}. The results are consistent if we instead use an alternative district definition that is constructed to facilitate the harmonization of census data over time \parencite{liu2023climate}. In practice, this implies that districts that later merged are combined into one consistent unit. These harmonized districts are the units of analysis for Table \ref{table:pca} and Panels A, B in \ref{table:balance_early}. We replicate our main results with the harmonized districts as the units of analysis, weighting tariff exposure and water pollution by district area. Using the harmonized districts as the unit of analysis yields similar, slightly larger point estimates.

\paragraph{Ban on leather tanneries} One reform that did effectively reduce water pollution was the ban of leather tanneries in Kanpur in 1987 \parencite{do2018can}. Since this ban occurred at the start of our study period, before trade liberalization, it is unlikely to affect our results. Excluding the three affected districts in our sample yields very similar results to the main estimates. 

\paragraph{Population weighting} Previous studies sometimes weight by population \parencite[e.g.,][]{topalova2010factor,bombardini2020trade}, although weighting sacrifices efficiency if effects are homogeneous \parencite[see, e.g.][]{solon2015we}. Using district populations from the 1991 census to weight observations yields slightly larger and less precisely estimated effects.   

\section{Conclusion}
\label{conclusion}

This paper studies the effect of India's 1991 trade liberalization reform on water pollution in rivers. Districts exposed to larger reductions in tariffs saw substantial increases in water pollution, compared to less-affected districts. We observe consistently-signed effects for six out of seven pollutants and one measure of water quality, as well as a summary pollution index. While comparing more- and less-exposed districts does not capture national-scale effects, our results seem primarily driven by increased pollution in more-exposed districts rather than systematic changes in pollution in less-exposed districts. The results are robust to a range of approaches to measurement and estimation. Although districts exposed to greater tariff reductions enacted water pollution abatement policies, we find no evidence that such policies were effective, consistent with \textcite{greenstone2014environmental}. 

We estimate the aggregate impacts of openness to trade at a district level. The main advantage of this approach is that we measure the overall effects of trade liberalization regardless of the channel through which effects operate. A limitation of this approach is that it does not provide information about how the effects arise. However, district-level census data suggest industrialization and structural transformation as potential mechanisms. 

Our analysis complements a literature that decomposes changes in firm-level emissions into scale, industry, and composition effects \parencite[e.g.,][]{barrows2021foreign,martin2011energy}. This literature primarily focuses on air pollution rather than water pollution. Bridging this gap may be an interesting avenue for future research, although firm-level data on water pollution remains extremely sparse \parencite{he2020watering}.

Our results add to a small but growing body of literature that shows that increased exposure to trade worsened environmental quality in low- and middle-income countries \parencite[e.g.,][]{bombardini2020trade,sekhri2022agricultural}. Within this literature, evidence from the Indian trade liberalization reform is an important complement to previous evidence because the reform was externally imposed and unexpected, and the changes in exposure to trade were uncorrelated with previous regional or industry-level trends \parencite{topalova2007,goldberg2010multiproduct,topalova2011trade}. This increases our confidence in our ability to distinguish the effects of trade liberalization from other, unrelated trends. 

Any negative environmental consequences of openness to trade must be weighed against the potential economic gains. Quantifying these environmental costs may, however, help policymakers to make more informed decisions about trade policy and design effective abatement and mitigation strategies.

\singlespacing

\clearpage

\printbibliography

\clearpage

\section*{Appendix: Trade and Pollution - Evidence from India}

\input{appendix.tex}

\end{document}

%% file: tables/table1.tex
{
\def\sym#1{\ifmmode^{#1}\else\(^{#1}\)\fi}
\begin{tabular}{l*{9}{c}}
\toprule
                &\multicolumn{1}{c}{(1)}&\multicolumn{1}{c}{(2)}&\multicolumn{1}{c}{(3)}&\multicolumn{1}{c}{(4)}&\multicolumn{1}{c}{(5)}&\multicolumn{1}{c}{(6)}&\multicolumn{1}{c}{(7)}&\multicolumn{1}{c}{(8)}&\multicolumn{1}{c}{(9)}\\
                &\multicolumn{1}{c}{Index}&\multicolumn{1}{c}{BOD}&\multicolumn{1}{c}{Chloride}&\multicolumn{1}{c}{DO}&\multicolumn{1}{c}{Hardness}&\multicolumn{1}{c}{Sulfate}&\multicolumn{1}{c}{TDS}&\multicolumn{1}{c}{Tot. coliform}&\multicolumn{1}{c}{Turbidity}\\
\midrule
Tariff          &    -2.73\sym{**} &    -3.24\sym{**} &    -4.83\sym{**} &    -3.47\sym{*}  &     0.32         &    -7.04\sym{**} &    -2.29         &    -1.94         &    -2.86\sym{**} \\
                &   (1.05)         &   (1.35)         &   (2.39)         &   (1.99)         &   (1.87)         &   (3.06)         &   (2.43)         &   (1.81)         &   (1.41)         \\
\midrule
N (observations)&     1146         &     1130         &     1136         &     1140         &     1134         &     1102         &      929         &      992         &     1062         \\
N (districts)   &      117         &      117         &      117         &      117         &      117         &      117         &      113         &      108         &      114         \\
\bottomrule
\end{tabular}
}

%% file: tables/table2.tex
{
\def\sym#1{\ifmmode^{#1}\else\(^{#1}\)\fi}
\begin{tabular}{l*{4}{c}}
\toprule
                &\multicolumn{1}{c}{(1)}&\multicolumn{1}{c}{(2)}&\multicolumn{1}{c}{(3)}&\multicolumn{1}{c}{(4)}\\
                &\multicolumn{1}{c}{Policy in place}&\multicolumn{1}{c}{Index}&\multicolumn{1}{c}{Index}&\multicolumn{1}{c}{Index}\\
\midrule
Tariff          &    -2.68\sym{*}  &    -2.73\sym{**} &    -2.81\sym{**} &    -3.29\sym{**} \\
                &   (1.36)         &   (1.05)         &   (1.09)         &   (1.33)         \\
\addlinespace
Policy in place&                  &                  &    -0.04         &                  \\
                &                  &                  &   (0.08)         &                  \\
\addlinespace
Policy in place $\times$ Tariff&                  &                  &     0.19         &                  \\
                &                  &                  &   (1.59)         &                  \\
\addlinespace
Ever implements policy $\times$ Tariff&                  &                  &                  &     0.87         \\
                &                  &                  &                  &   (1.40)         \\
\midrule
N (observations)&     1146         &     1146         &     1146         &     1146         \\
N (districts)   &      117         &      117         &      117         &      117         \\
\bottomrule
\end{tabular}
}

%% file: appendix.tex
\setcounter{table}{0}
\setcounter{figure}{0}
\renewcommand{\thesubsection}{A\arabic{subsection}}
\renewcommand{\thetable}{A\arabic{table}}
\renewcommand{\thefigure}{A\arabic{figure}}

\subsection{Water quality metrics}
\label{app:pollutants}

Below follows a description of the water quality metrics we use to measure water pollution in this paper, and the units in which they are measured. We focus primarily on seven pollutants and one measure of water quality. The seven pollutants are:

\paragraph{BOD (mg/l)} Biochemical Oxygen Demand (BOD) is an important and commonly used indicator of water quality that quantifies the presence of organic matter. BOD measures how much oxygen is used by microorganisms in the sample to break down organic waste over a fixed period \parencite{chapman1996water}. Higher BOD indicates more organic pollution, typically originating from farming, food processing, wastewater discharge, or domestic sewage \parencite{chapman2013surface}. The breakdown of organic pollution can deplete oxygen in water bodies, harming aquatic life and ecosystem health \parencite{chapman2013surface}.

\paragraph{Chloride (mg/l)} Chloride is an anion that occurs naturally in the environment. Increasing concentrations are at least partly attributable to anthropogenic sources, including wastewater, fertilizer, animal waste, irrigation, aquaculture, energy production wastes, and landfill leachates \parencite{granato2015methods}. Chloride combines with cations such as sodium or calcium to form salts, thereby increasing salinity. High salinity can harm aquatic plants and animals \parencite{carr2008water} and can render water unfit for human or livestock consumption \parencite{chapman1996water}. 

\paragraph{Hardness (mg/l)} Hardness measures the mineral content of water, primarily the concentration of calcium and magnesium salts \parencite{chapman1996water}. Hardness in surface water is primarily influenced by the geology of the catchment area \parencite{WHO2011hardness}. Higher levels indicate harder water \parencite{carr2008water}. Hardness is not directly harmful to human or aquatic life, but can result in scale deposits in pipes \parencite{WHO2011hardness}.

\paragraph{Turbidity (NTU)} Turbidity  measures the transparency of water. Higher values mean lower transparency, driven by the presence of suspended particles, including clay, silt, other sedimentary particles, and organic detritus \parencite{carr2008water}. Human activity upstream in the watershed can increase sedimentation and turbidity downstream, for example through changes in soil erosion patterns, urban runoff, or the release of industrial effluents or nutrients that increase the growth of phytoplankton \parencite{carr2008water}. Because turbidity reduces light penetration, high turbidity affects the growth of aquatic plants and reduces the ability of visual predators to hunt. Pathogenic bacteria may be present in highly turbid water \parencite{carr2008water}.

\paragraph{TDS (mg/l)} Total Dissolved Solids (TDS) measures the quantity of solids in water that can pass through a very fine filter, typically comprising inorganic salts and small amounts of organic matter \parencite{world2003tds}. Anthropogenic sources include sewage, urban and agricultural run-off, and industrial wastewater \parencite{world2003tds}. The health and environmental impacts of high TDS depend on the source, but high TDS is often an indicator of the presence of contaminants \parencite{usgs_chloride_salinity_dissolved2019}. 

\paragraph{Total coliforms (MPN/100 ml)} Total coliforms measures the most probable number (MPN) of coliform bacteria per 100 ml of water. Coliform bacteria are a large group of bacteria that are common in the natural environment. Most coliform bacteria are not harmful to human health.  However, a subset of these bacteria, fecal coliforms, inhabit the intestines of warm-blooded animals. The presence of coliform bacteria can thus indicate contamination with human or animal fecal waste. 

\paragraph{Sulfate (mg/l)} Sulfate is an anion that is present in all natural waters, but higher concentrations indicate contamination. Major anthropogenic sources of sulfate include industrial atmospheric pollution, mines, fertiliser use, and industrial wastewater discharge, in particular from pulp and paper mills, textile mills and tanneries \parencite{zak2021sulphate,WHO2004_sulfate}. There is limited information about the human health effects of exposure to sulfates, although high levels of sulfates can cause diarrhea \parencite{zak2021sulphate, WHO2004_sulfate}. Sulfates themselves are toxic to aquatic life at high concentrations \parencite{zak2021sulphate}. Also, in low-oxygen environments, organisms can metabolize sulfate to produce sulfide, which is toxic to many aquatic life forms \parencite{zak2021sulphate,lamers2013sulfide}.  

\noindent \paragraph{} We also report effects on one measure of water quality:

\paragraph{DO (mg/l)} Dissolved Oxygen (DO) measures the amount of oxygen dissolved in water, which is critical for the survival of aquatic life. Low levels of DO suggest high levels of pollution with organic waste, with similar potential sources as for high BOD \parencite{chapman1996water}. High levels of DO suggest good water quality \parencite{carr2008water}.

\subsection{Inverse-covariance weighted index with missing components}
\label{app:anderson_index}

Suppose we observe $J$ pollution metrics for district $i$ in year $t$, collected in the vector
$X_{it} = (X_{it1},\dots,X_{itJ})'$.
Let $S_{it} \in \{0,1\}$ indicate that $t <$ 1991, i.e., before the reform period. 

\paragraph{Step 1: Standardize components using pre-reform observations.}
For each pollution metric $j$, compute the pre-reform mean and standard deviation:
\begin{align*}
\hat\mu_j &= \frac{\sum_i S_{it} X_{itj}}{\sum_i S_{it}},\\
\hat\sigma_j &= \sqrt{\frac{\sum_i S_{it} (X_{itj}-\hat\mu_j)^2}{\sum_i S_{it}}}.
\end{align*}
Define the standardized component:
\begin{equation*}
Z_{itj} = \frac{X_{itj}-\hat\mu_j}{\hat\sigma_j}.
\end{equation*}

\paragraph{Step 2: Compute inverse-covariance (inverse-correlation) weights.}
Let $\hat R$ denote the $J\times J$ pairwise correlation matrix of the pollution metrics
$\{X_{itj}\}_{j=1}^J$, constructed using all available (non-missing) observations for each pair.
Let $\hat R^{-1}$ be its inverse.
Following \textcite{anderson2008multiple}, define the raw index weight for pollution metric $j$ as the row-sum of the inverse matrix:
\begin{equation*}
\tilde w_j = \sum_{k=1}^J (\hat R^{-1})_{jk},
\qquad j=1,\dots,J.
\end{equation*}

\paragraph{Step 3: Adjust for missing components at the district-year level.}
Let $M_{itj} = \mathbbm{1}\{X_{itj}\ \text{is observed}\}$ indicate whether component $j$ is non-missing for district $i$ in year $t$.
The procedure sets the contribution of missing components to zero and re-normalizes weights over the set of observed components for each unit:
\begin{align*}
W_{it} &= \sum_{j=1}^J M_{itj}\tilde w_j,\\
\omega_{itj} &=
\begin{cases}
\dfrac{M_{itj}\tilde w_j}{W_{it}}, & \text{if } W_{it}>0,\\[6pt]
\text{undefined}, & \text{if } W_{it}=0.
\end{cases}
\end{align*}

\paragraph{Step 4: Construct the index.}
The inverse-covariance weighted index is the weighted average of standardized components:
\begin{equation*}
I_{it} = \sum_{j=1}^J \omega_{itj} Z_{itj},
\end{equation*}
with the index $I_{it}$ missing when $W_{it}=0$ (all components missing).

\clearpage


\begin{figure}[htbp!]
    \centering
    \includegraphics[width=1\textwidth]{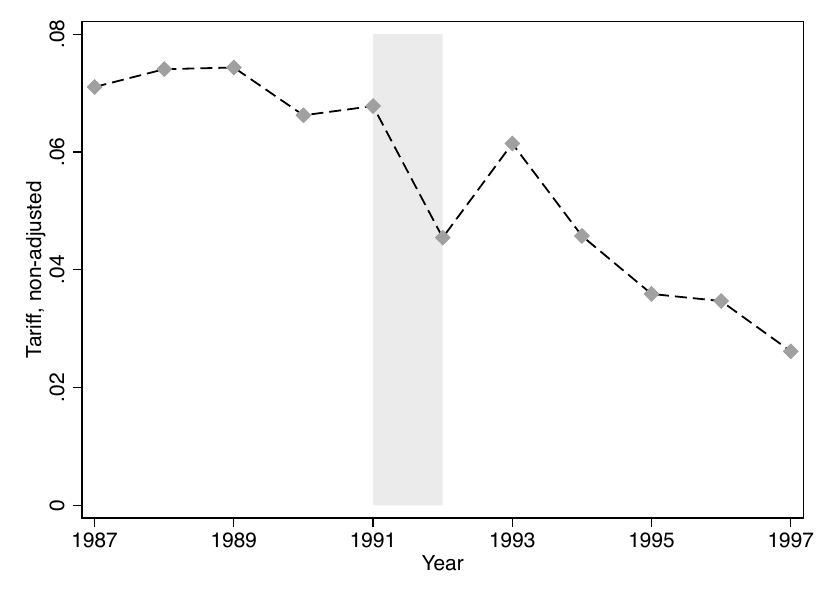}
    \caption{\raggedright Average tariff exposure for districts in our sample, based on raw data, without interpolating values for the year 1993. Shaded area indicates the reform period.}
    \label{fig:trend_tariff_nonadj}
\end{figure}

\begin{figure}[htbp!]
    \centering
    \includegraphics[width=0.8\textwidth]{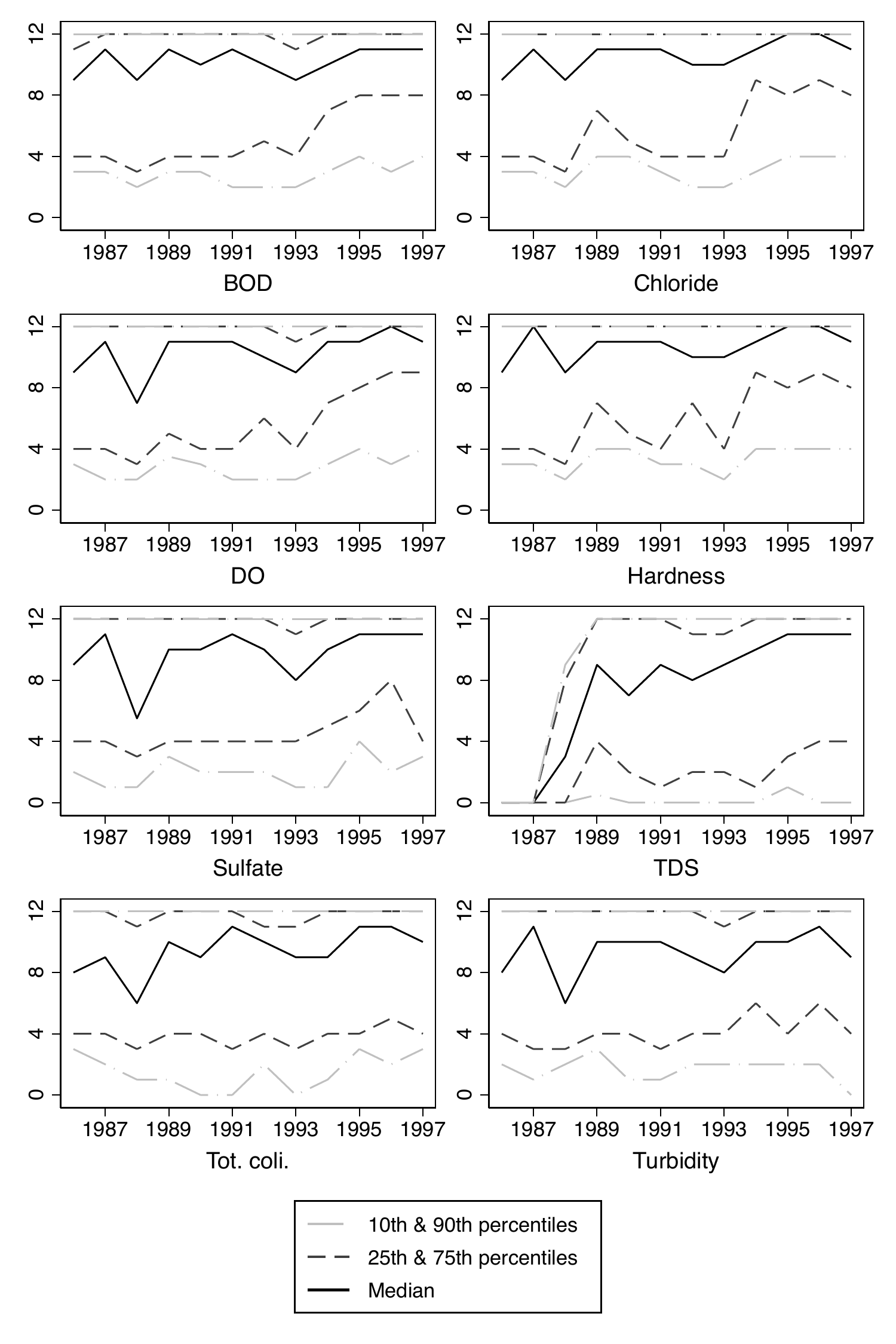}
    \caption{\raggedright The number of observations for each pollutant at the station-year level between 1986 and 1997. The number of observations varies both because of the underlying cadence of monitoring and because of missing data: 12 observations is equivalent to complete monthly data; 4 observations is equivalent to complete quarterly data.}
    \label{fig:pollution_count}
\end{figure}

\begin{figure}[htbp!]
    \centering
    \includegraphics[width=0.9\textwidth]{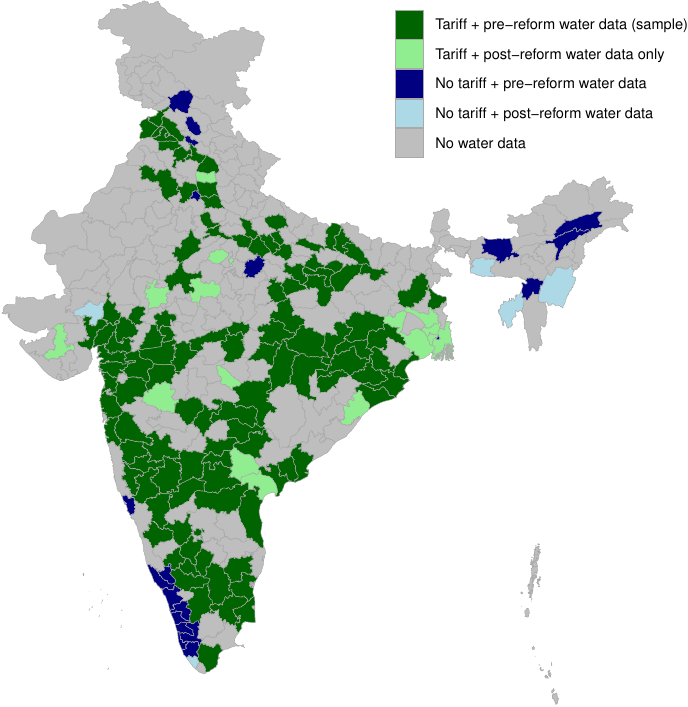}
    \caption{\raggedright Sample construction. Districts with both tariff data and pre-reform water pollution data: 117 (main sample). Districts with pre-reform water pollution data, excluded because of missing tariff data: 20. Districts with post-reform water pollution data, excluded because of missing pre-reform water pollution data: 21 (16 with non-missing tariff data, 5 with missing tariff data). All other districts have no reported water quality data. Data sources as described in the main text.}
    \label{fig:water_map}
\end{figure}

\begin{figure}[htbp!]
  \centering
  \begin{subfigure}{0.8\textwidth}
    \includegraphics[width=\linewidth]{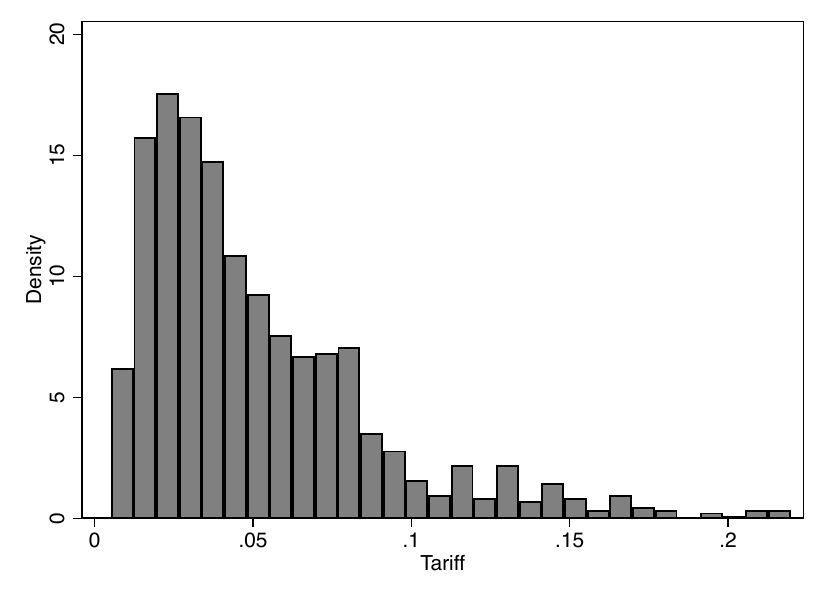}
    \caption{}
    \label{fig:hist_tariff_y}
  \end{subfigure}
  \begin{subfigure}{0.8\textwidth}
    \includegraphics[width=\linewidth]{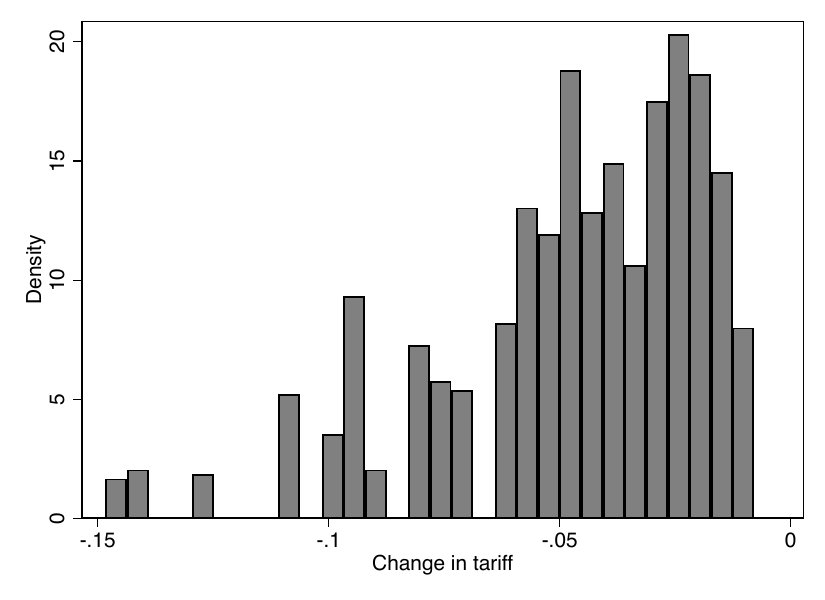}
    \caption{}
    \label{fig:hist_tariff_Dy}
  \end{subfigure}

  \caption{\raggedright Panel a) shows the distribution of the tariff measure values across observation-years for all districts in our sample. Panel b) shows the distribution of the tariff measure changes within districts between 1987 and 1997.}
  
\end{figure}

\begin{figure}[htbp!]
    \centering
    \includegraphics[width=1\textwidth]{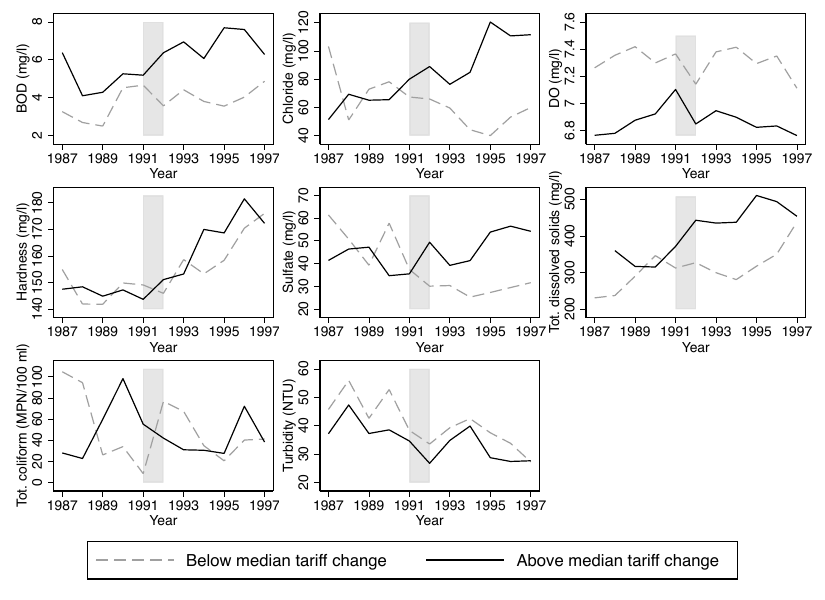}
    \caption{\raggedright The figure shows the trend of each pollutant by groups of districts, depending on whether a district is above or below median tariff change. The shaded area shows the period when tariffs started falling.}
    \label{fig:trend_all_pollutants}
\end{figure}

\begin{figure}[htbp!]
    \centering
    \includegraphics[width=1\textwidth]{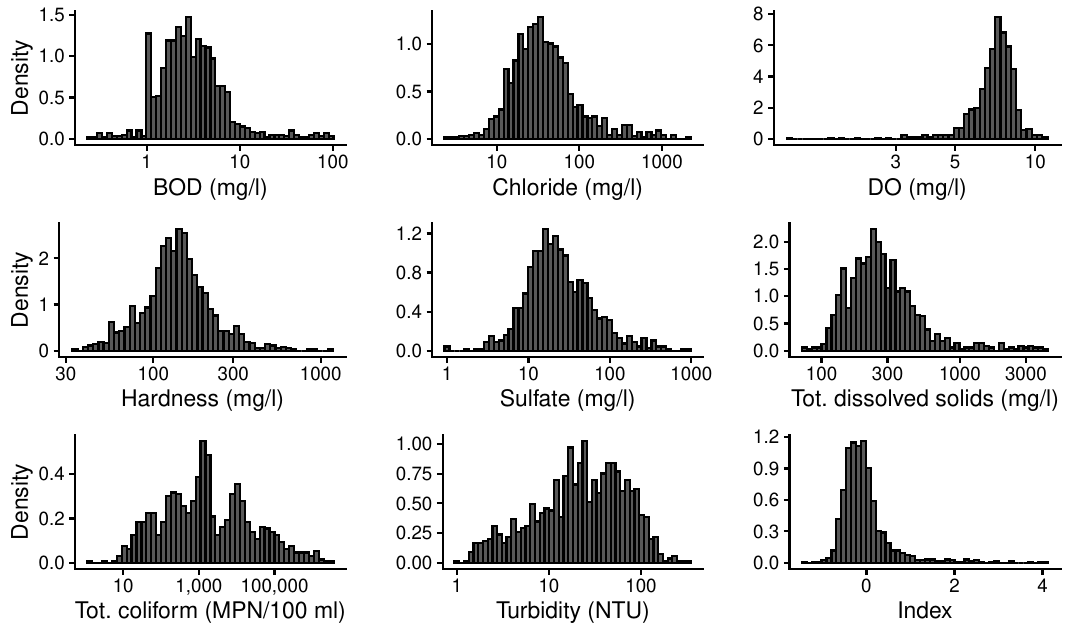}
    \caption{\raggedright Histograms show the individual distribution of each water quality metric and the water pollution index, where each district-year observation in our sample is included. The horizontal axes are on a logarithmic scale.}
    \label{fig:hist_water}
\end{figure}

\begin{figure}[t!]
  \centering

  \begin{subfigure}{0.8\textwidth}
    \includegraphics[width=\linewidth]{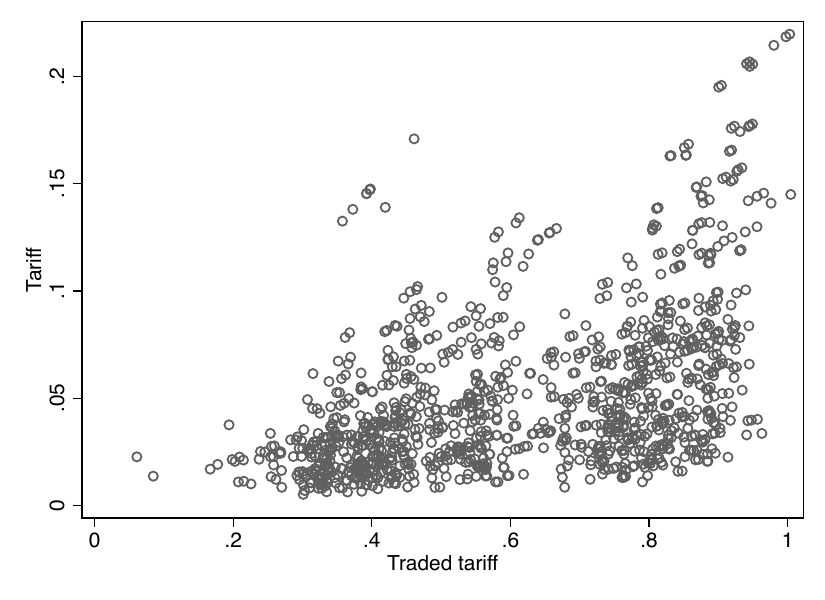}
    \caption{Raw data}
    \label{fig:fs_raw}
  \end{subfigure}
  \begin{subfigure}{0.8\textwidth}
    \includegraphics[width=\linewidth]{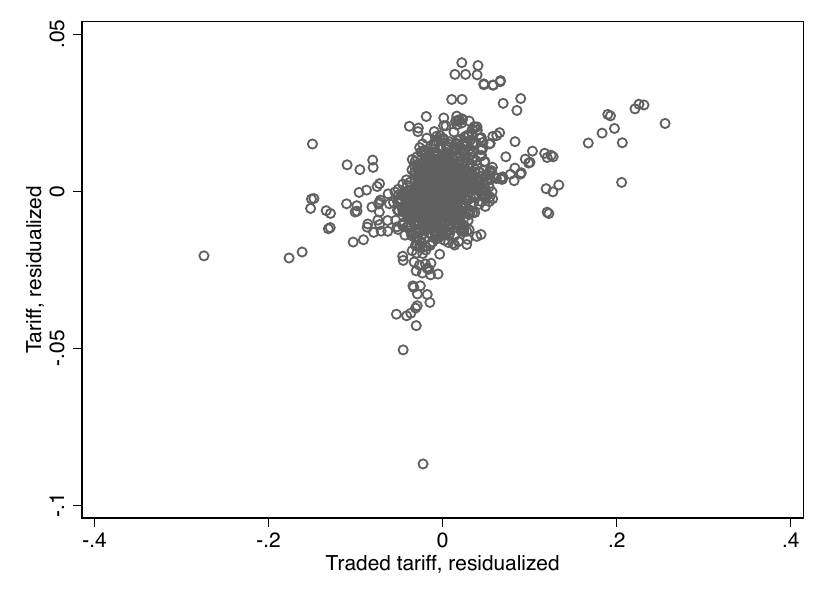}
    \caption{Residualized data}
    \label{fig:fs_res}
  \end{subfigure}

  \caption{\raggedright Graphs visualize the first stage relationship between tariff exposure and traded tariff exposure. Panel a) plots the relationship in the raw data between tariff exposure and traded tariff exposure. Panel b) plots the same relationship after residualizing with respect to district and year fixed effects.\label{fig:fs}}
  
\end{figure}

\clearpage

\begin{figure}[t!]
  \centering
    \includegraphics[width=\linewidth]{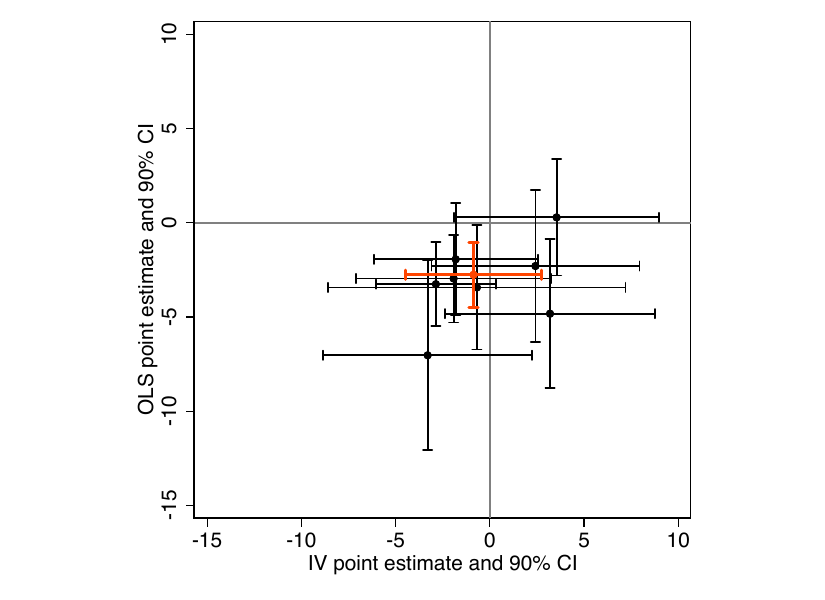}
  \caption{\raggedright Graph plots main OLS estimates against IV estimates that use traded tariff exposure as an instrument for tariff exposure. Black dots show point estimates for individual pollutant metrics. Red dot shows point estimates for pollution index. Whiskers show 90\% confidence intervals.  All regressions include year and district fixed effects. Standard errors clustered by district.    \label{fig:ols_iv}}
  
\end{figure}

\begin{figure}[htbp!]
    \centering
    \includegraphics[width=1\textwidth]{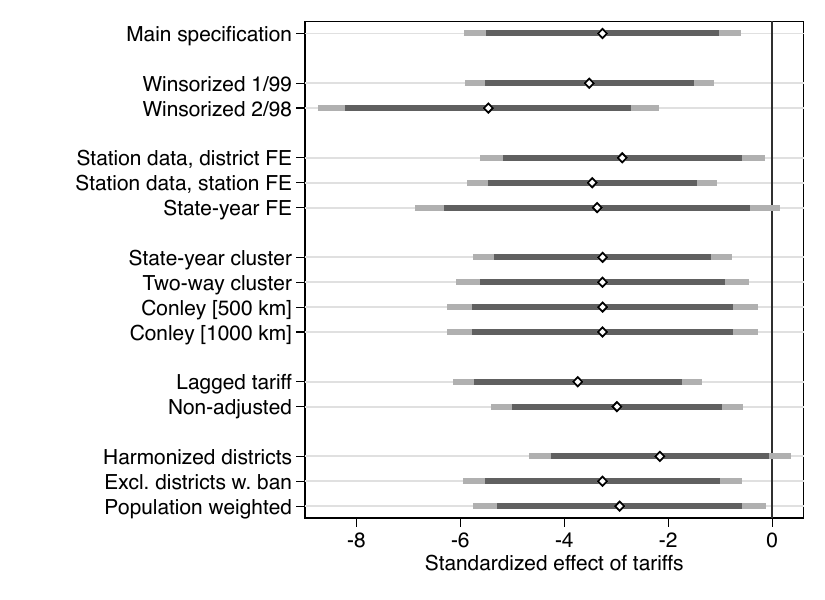}
    \caption{\raggedright \textbf{Robustness of effect on BOD}. Main specification: standardized BOD concentrations, contemporaneous effects, district and year fixed effects, standard errors clustered by district. Robustness tests vary analysis choices, as follows: Winsorized: extreme 2 or 4 percentiles of pollution data winsorized. Station data, district FE: main specification using monitor station-level data. Station data, station FE: monitoring-station fixed effects. State-year FE: global trends modeled with state-year fixed effects. State-year cluster: standard errors clustered by state-year. Two-way cluster: standard errors clustered two-way by district and year. Conley: Conley standard errors in addition to clustering by district, over 500 and 1000 km respectively. Lagged: effects of tariffs lagged one year. Non-adjusted: tariff data without adjusting value in 1993. Harmonized districts: unit of analysis is a harmonized district  \parencite{liu2023climate}, with area-weighted tariff and pollution variables. Excl. districts w. ban: drops 3 districts affected by a local ban on leather tanneries. Population weighted: districts weighted by 1991 population. Dark grey confidence intervals at the 90\% level; lighter grey confidence intervals at the 95\% level. N = 1,130 observations (2,452 at the station level) in 117 districts, except with lagged tariffs (N = 1,1015 in 117 districts), in harmonized district sample (N = 883 in 91 districts), and excluding districts with the tannery ban (N = 1,101 in 114 districts).}
    \label{fig:robustness_bod}
\end{figure}

\begin{figure}[htbp!]
    \centering
    \includegraphics[width=1\textwidth]{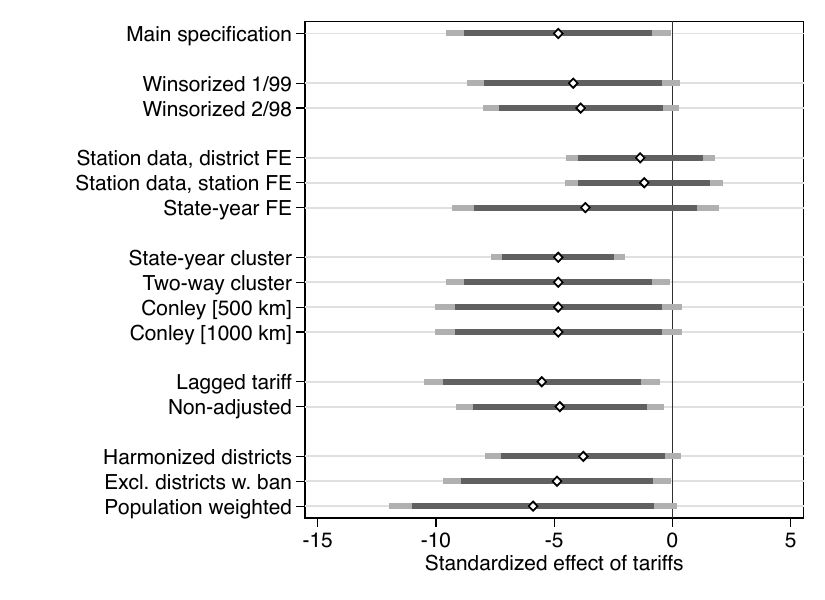}
    \caption{\raggedright \textbf{Robustness of effect on chloride}. Main specification: standardized chloride concentrations, contemporaneous effects, district and year fixed effects, standard errors clustered by district. Robustness tests vary analysis choices, as follows: Winsorized: extreme 2 or 4 percentiles of pollution data winsorized. Station data, district FE: main specification using monitor station-level data. Station data, station FE: monitoring-station fixed effects. State-year FE: global trends modeled with state-year fixed effects. State-year cluster: standard errors clustered by state-year. Two-way cluster: standard errors clustered two-way by district and year. Conley: Conley standard errors in addition to clustering by district, over 500 and 1000 km respectively. Lagged: effects of tariffs lagged one year. Non-adjusted: tariff data without adjusting value in 1993. Harmonized districts: unit of analysis is a harmonized district  \parencite{liu2023climate}, with area-weighted tariff and pollution variables. Excl. districts w. ban: drops 3 districts affected by a local ban on leather tanneries. Population weighted: districts weighted by 1991 population. Dark grey confidence intervals at the 90\% level; lighter grey confidence intervals at the 95\% level. N = 1,136 observations (2,467 at the station level) in 117 districts, except with lagged tariffs (N = 1,020 in 117 districts), in harmonized district sample (N = 889 in 91 districts), and excluding districts with the tannery ban (N = 1,107 in 114 districts).}
    \label{fig:robustness_chlor}
\end{figure}

\begin{figure}[htbp!]
    \centering
    \includegraphics[width=1\textwidth]{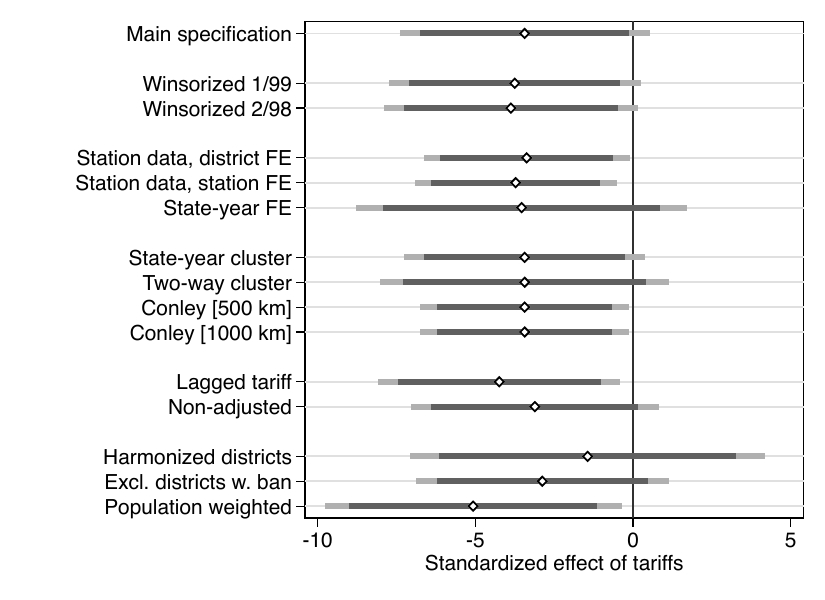}
    \caption{\raggedright \textbf{Robustness of effect on DO}. Main specification: standardized DO concentrations, contemporaneous effects, district and year fixed effects, standard errors clustered by district. Robustness tests vary analysis choices, as follows: Winsorized: extreme 2 or 4 percentiles of pollution data winsorized. Station data, district FE: main specification using monitor station-level data. Station data, station FE: monitoring-station fixed effects. State-year FE: global trends modeled with state-year fixed effects. State-year cluster: standard errors clustered by state-year. Two-way cluster: standard errors clustered two-way by district and year. Conley: Conley standard errors in addition to clustering by district, over 500 and 1000 km respectively. Lagged: effects of tariffs lagged one year. Non-adjusted: tariff data without adjusting value in 1993. Harmonized districts: unit of analysis is a harmonized district  \parencite{liu2023climate}, with area-weighted tariff and pollution variables. Excl. districts w. ban: drops 3 districts affected by a local ban on leather tanneries. Population weighted: districts weighted by 1991 population. Dark grey confidence intervals at the 90\% level; lighter grey confidence intervals at the 95\% level. N = 1,140 observations (2,470 at the station level) in 117 districts, except with lagged tariffs (N = 1,027 in 117 districts), in harmonized district sample (N = 891 in 91 districts), and excluding districts with the tannery ban (N = 1,111 in 114 districts).}
    \label{fig:robustness_do}
\end{figure}

\begin{figure}[htbp!]
    \centering
    \includegraphics[width=1\textwidth]{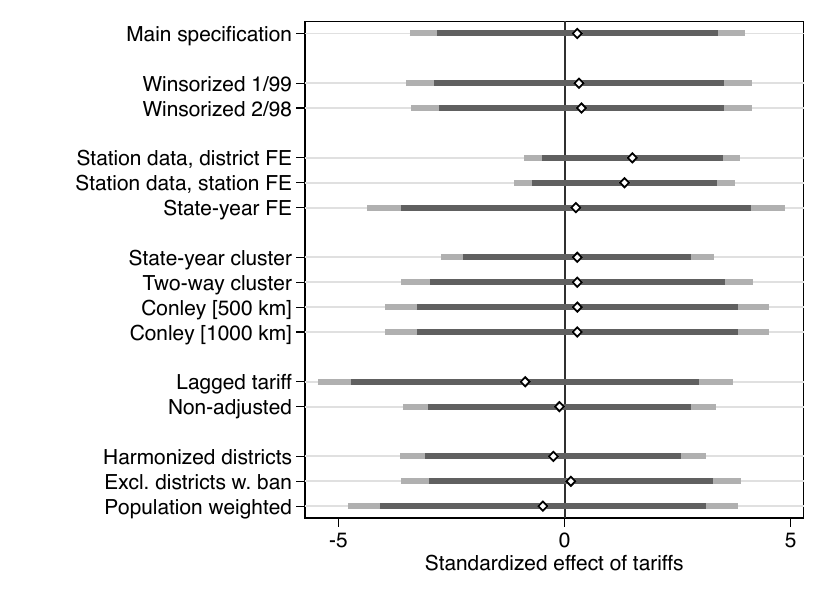}
    \caption{\raggedright \textbf{Robustness of effect on hardness}. Main specification: standardized hardness concentrations, contemporaneous effects, district and year fixed effects, standard errors clustered by district. Robustness tests vary analysis choices, as follows: Winsorized: extreme 2 or 4 percentiles of pollution data winsorized. Station data, district FE: main specification using monitor station-level data. Station data, station FE: monitoring-station fixed effects. State-year FE: global trends modeled with state-year fixed effects. State-year cluster: standard errors clustered by state-year. Two-way cluster: standard errors clustered two-way by district and year. Conley: Conley standard errors in addition to clustering by district, over 500 and 1000 km respectively. Lagged: effects of tariffs lagged one year. Non-adjusted: tariff data without adjusting value in 1993. Harmonized districts: unit of analysis is a harmonized district  \parencite{liu2023climate}, with area-weighted tariff and pollution variables. Excl. districts w. ban: drops 3 districts affected by a local ban on leather tanneries. Population weighted: districts weighted by 1991 population. Dark grey confidence intervals at the 90\% level; lighter grey confidence intervals at the 95\% level. N = 1,134 observations (2,464 at the station level) in 117 districts, except with lagged tariffs (N = 1,017 in 117 districts), in harmonized district sample (N = 889 in 91 districts), and excluding districts with the tannery ban (N = 1,105 in 114 districts).}
    \label{fig:robustness_hardness}
\end{figure}

\begin{figure}[htbp!]
    \centering
    \includegraphics[width=1\textwidth]{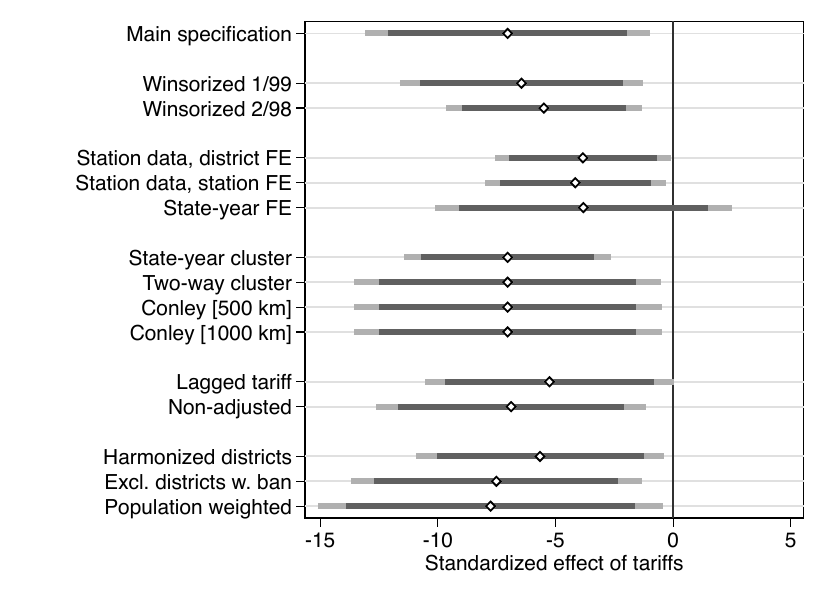}
    \caption{\raggedright \textbf{Robustness of effect on sulfate}. Main specification: standardized sulfate concentrations, contemporaneous effects, district and year fixed effects, standard errors clustered by district. Robustness tests vary analysis choices, as follows: Winsorized: extreme 2 or 4 percentiles of pollution data winsorized. Station data, district FE: main specification using monitor station-level data. Station data, station FE: monitoring-station fixed effects. State-year FE: global trends modeled with state-year fixed effects. State-year cluster: standard errors clustered by state-year. Two-way cluster: standard errors clustered two-way by district and year. Conley: Conley standard errors in addition to clustering by district, over 500 and 1000 km respectively. Lagged: effects of tariffs lagged one year. Non-adjusted: tariff data without adjusting value in 1993. Harmonized districts: unit of analysis is a harmonized district  \parencite{liu2023climate}, with area-weighted tariff and pollution variables. Excl. districts w. ban: drops 3 districts affected by a local ban on leather tanneries. Population weighted: districts weighted by 1991 population. Dark grey confidence intervals at the 90\% level; lighter grey confidence intervals at the 95\% level. N = 1,102 observations (2,406 at the station level) in 117 districts, except with lagged tariffs (N = 994 in 117 districts), in harmonized district sample (N = 860 in 91 districts), and excluding districts with the tannery ban (N = 1,076 in 114 districts).}
    \label{fig:robustness_sulph}
\end{figure}

\begin{figure}[htbp!]
    \centering
    \includegraphics[width=1\textwidth]{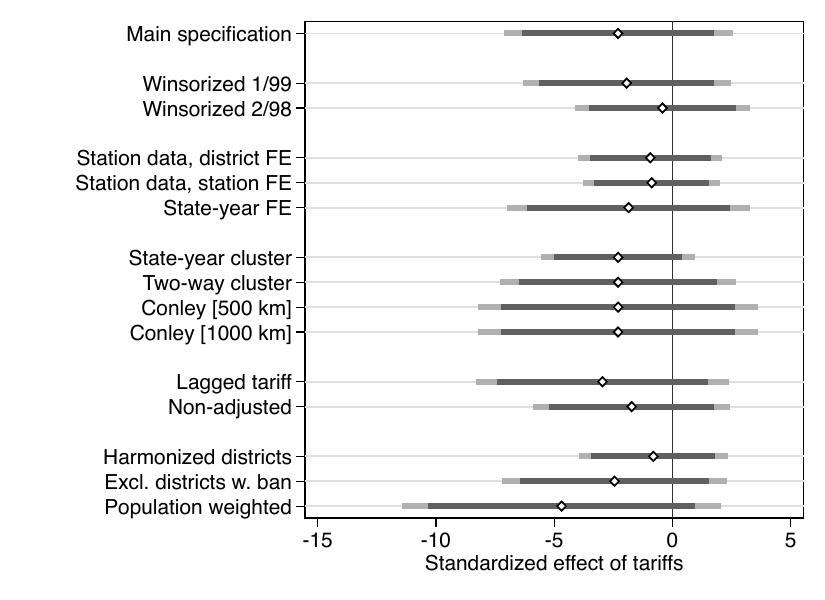}
    \caption{\raggedright \textbf{Robustness of effect on total dissolved solids (TDS)}. Main specification: standardized TDS concentrations, contemporaneous effects, district and year fixed effects, standard errors clustered by district. Robustness tests vary analysis choices, as follows: Winsorized: extreme 2 or 4 percentiles of pollution data winsorized. Station data, district FE: main specification using monitor station-level data. Station data, station FE: monitoring-station fixed effects. State-year FE: global trends modeled with state-year fixed effects. State-year cluster: standard errors clustered by state-year. Two-way cluster: standard errors clustered two-way by district and year. Conley: Conley standard errors in addition to clustering by district, over 500 and 1000 km respectively. Lagged: effects of tariffs lagged one year. Non-adjusted: tariff data without adjusting value in 1993. Harmonized districts: unit of analysis is a harmonized district  \parencite{liu2023climate}, with area-weighted tariff and pollution variables. Excl. districts w. ban: drops 3 districts affected by a local ban on leather tanneries. Population weighted: districts weighted by 1991 population. Dark grey confidence intervals at the 90\% level; lighter grey confidence intervals at the 95\% level. N = 929 observations (1,988 at the station level) in 113 districts, except with lagged tariffs (N = 876 in 113 districts), in harmonized district sample (N = 724 in 87 districts), and excluding districts with the tannery ban (N = 904 in 110 districts).}
    \label{fig:robustness_tds}
\end{figure}

\begin{figure}[htbp!]
    \centering
    \includegraphics[width=1\textwidth]{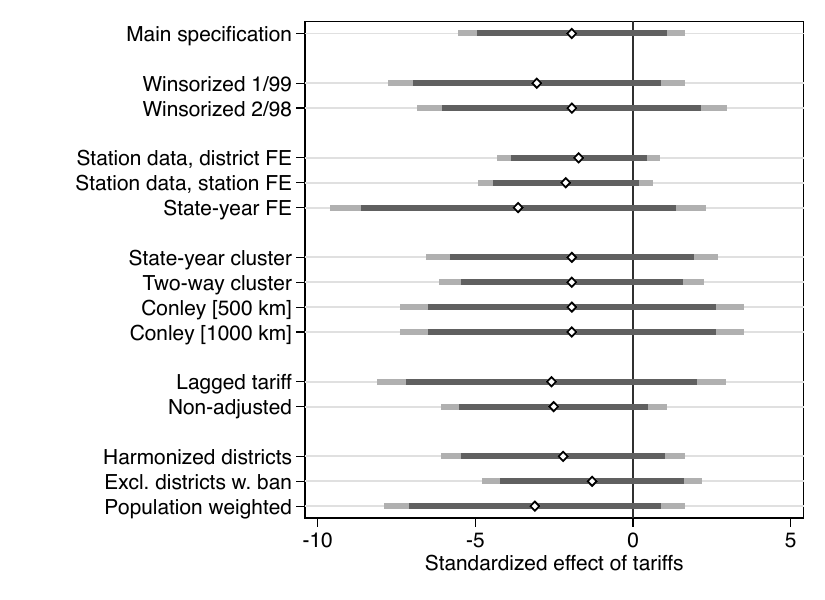}
    \caption{\raggedright \textbf{Robustness of effect on total coliform}. Main specification: standardized total coliform concentrations, contemporaneous effects, district and year fixed effects, standard errors clustered by district. Robustness tests vary analysis choices, as follows: Winsorized: extreme 2 or 4 percentiles of pollution data winsorized. Station data, district FE: main specification using monitor station-level data. Station data, station FE: monitoring-station fixed effects. State-year FE: global trends modeled with state-year fixed effects. State-year cluster: standard errors clustered by state-year. Two-way cluster: standard errors clustered two-way by district and year. Conley: Conley standard errors in addition to clustering by district, over 500 and 1000 km respectively. Lagged: effects of tariffs lagged one year. Non-adjusted: tariff data without adjusting value in 1993. Harmonized districts: unit of analysis is a harmonized district  \parencite{liu2023climate}, with area-weighted tariff and pollution variables. Excl. districts w. ban: drops 3 districts affected by a local ban on leather tanneries. Population weighted: districts weighted by 1991 population. Dark grey confidence intervals at the 90\% level; lighter grey confidence intervals at the 95\% level. N = 992 observations (2,256 at the station level) in 108 districts, except with lagged tariffs (N = 891 in 107 districts), in harmonized district sample (N = 781 in 84 districts), and excluding districts with the tannery ban (N = 966 in 105 districts).}
    \label{fig:robustness_totcoli}
\end{figure}

\begin{figure}[htbp!]
    \centering
    \includegraphics[width=1\textwidth]{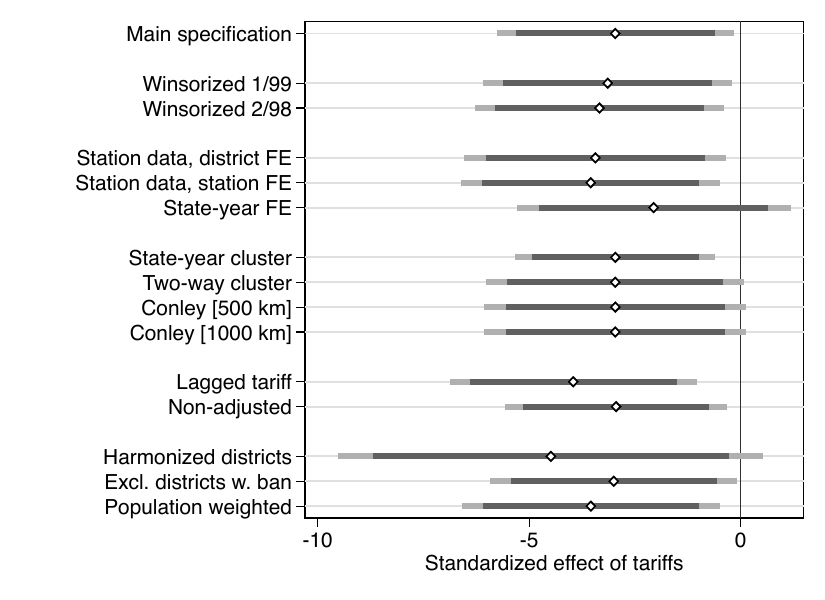}
    \caption{\raggedright \textbf{Robustness of effect on turbidity}. Main specification: standardized turbidity concentrations, contemporaneous effects, district and year fixed effects, standard errors clustered by district. Robustness tests vary analysis choices, as follows: Winsorized: extreme 2 or 4 percentiles of pollution data winsorized. Station data, district FE: main specification using monitor station-level data. Station data, station FE: monitoring-station fixed effects. State-year FE: global trends modeled with state-year fixed effects. State-year cluster: standard errors clustered by state-year. Two-way cluster: standard errors clustered two-way by district and year. Conley: Conley standard errors in addition to clustering by district, over 500 and 1000 km respectively. Lagged: effects of tariffs lagged one year. Non-adjusted: tariff data without adjusting value in 1993. Harmonized districts: unit of analysis is a harmonized district  \parencite{liu2023climate}, with area-weighted tariff and pollution variables. Excl. districts w. ban: drops 3 districts affected by a local ban on leather tanneries. Population weighted: districts weighted by 1991 population. Dark grey confidence intervals at the 90\% level; lighter grey confidence intervals at the 95\% level. N = 1,062 observations (2,336 at the station level) in 114 districts, except with lagged tariffs (N = 956 in 114 districts), in harmonized district sample (N = 839 in 89 districts), and excluding districts with the tannery ban (N = 1,033 in 111 districts).}
    \label{fig:robustness_turb}
\end{figure}

\begin{figure}[htbp!]
    
    \centering
    \includegraphics[width=1\textwidth]{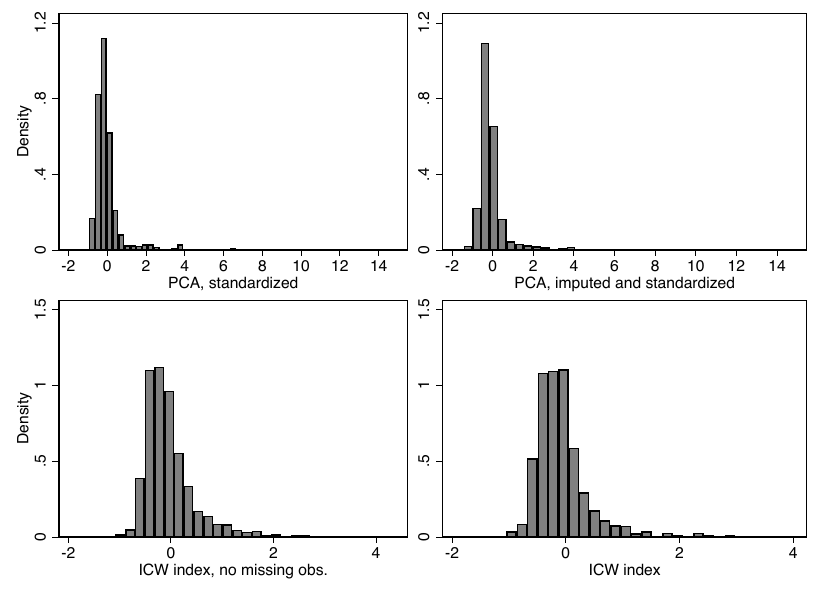}
    \caption{\raggedright %
     Distribution of water pollution index values, using principal component analysis (top row) and inverse-covariance weighting (bottom row). The left-hand column shows distributions for a sample of districts with no missing data for any pollutant. The right-hand column shows distributions for the main sample. For the PCA index, this implies imputing missing pollutant data.\label{fig:index_distribution_comp}}  
\end{figure}

\begin{figure}[htbp!]
    \centering
    \includegraphics[width=1\textwidth]{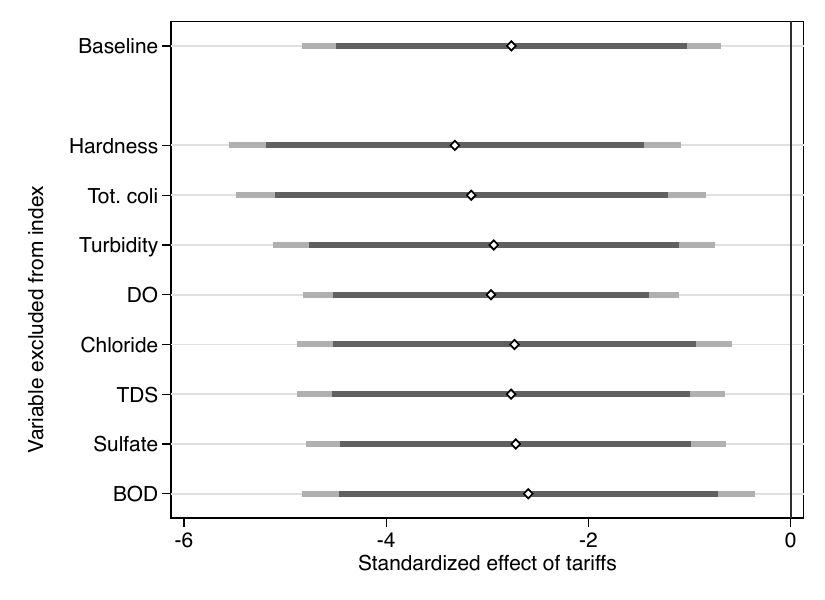}
    \caption{\raggedright Results from varying which variables are included in constructing the water pollution index. Each estimate excludes the labeled variable in constructing the metric. All estimates show the standardized effect of a contemporaneous change in the tariff measure. District and year fixed effects are included in all regressions. Standard errors are clustered at the district level. The confidence intervals colored dark gray are at the 90\% level, while lighter gray confidence intervals are at the 95\% level.}
    \label{fig:coefplot_loo}
\end{figure}

\begin{figure}[htbp!]
  \centering

  \begin{subfigure}{0.8\textwidth}
    \includegraphics[width=\linewidth]{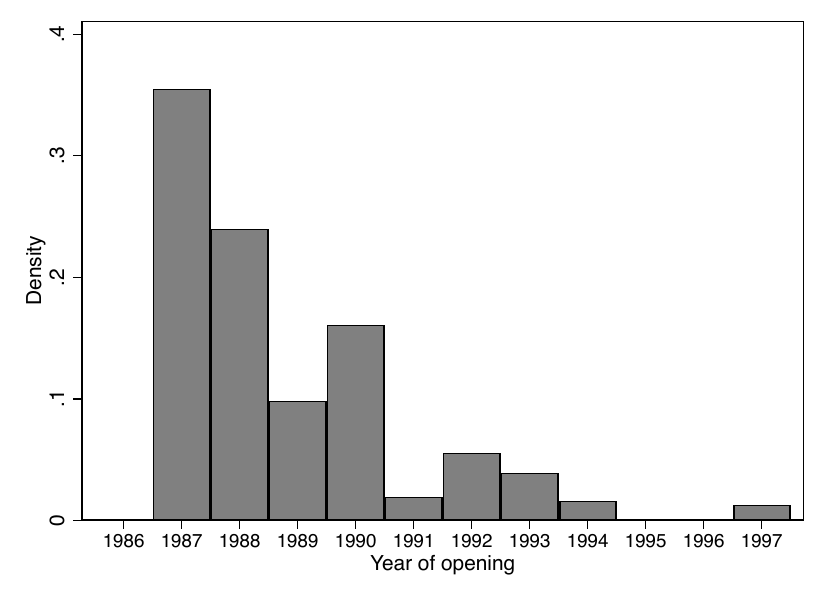}
    \caption{}
    \label{fig:hist_yop}
  \end{subfigure}
  \begin{subfigure}{0.8\textwidth}
    \includegraphics[width=\linewidth]{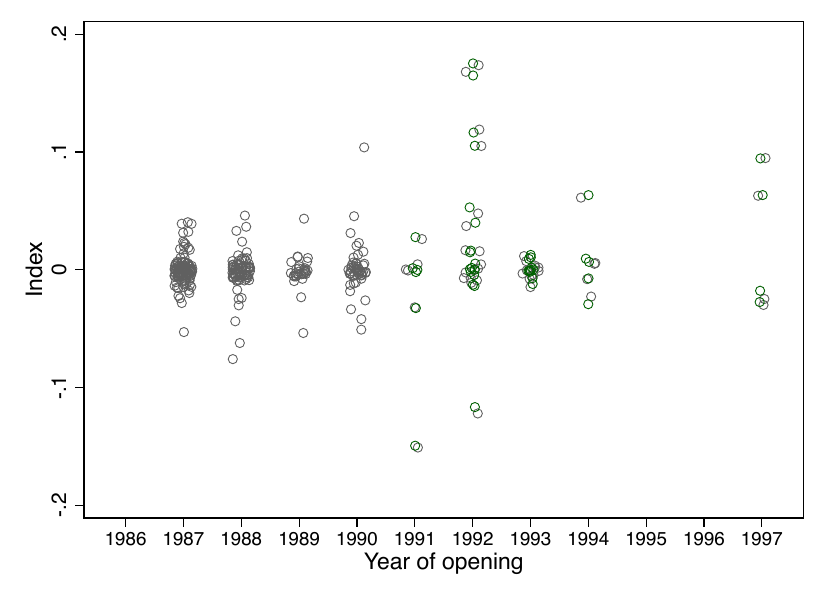}
    \caption{}
    \label{fig:index_res_yop}
  \end{subfigure}

  \caption{\raggedright Panel a) shows the distribution of year of opening of monitoring stations in our sample. Panel b) shows the residualized (in a regression with district and year fixed effects) values of the water index plotted against the year in which a  monitoring station opened in the district. Stations opened before 1991 are colored gray; stations opened in 1991 or after are colored green.\label{fig:yop}}
  
\end{figure}

\begin{figure}[htbp!]
  \centering

  \begin{subfigure}{1\textwidth}
    \includegraphics[width=\linewidth]{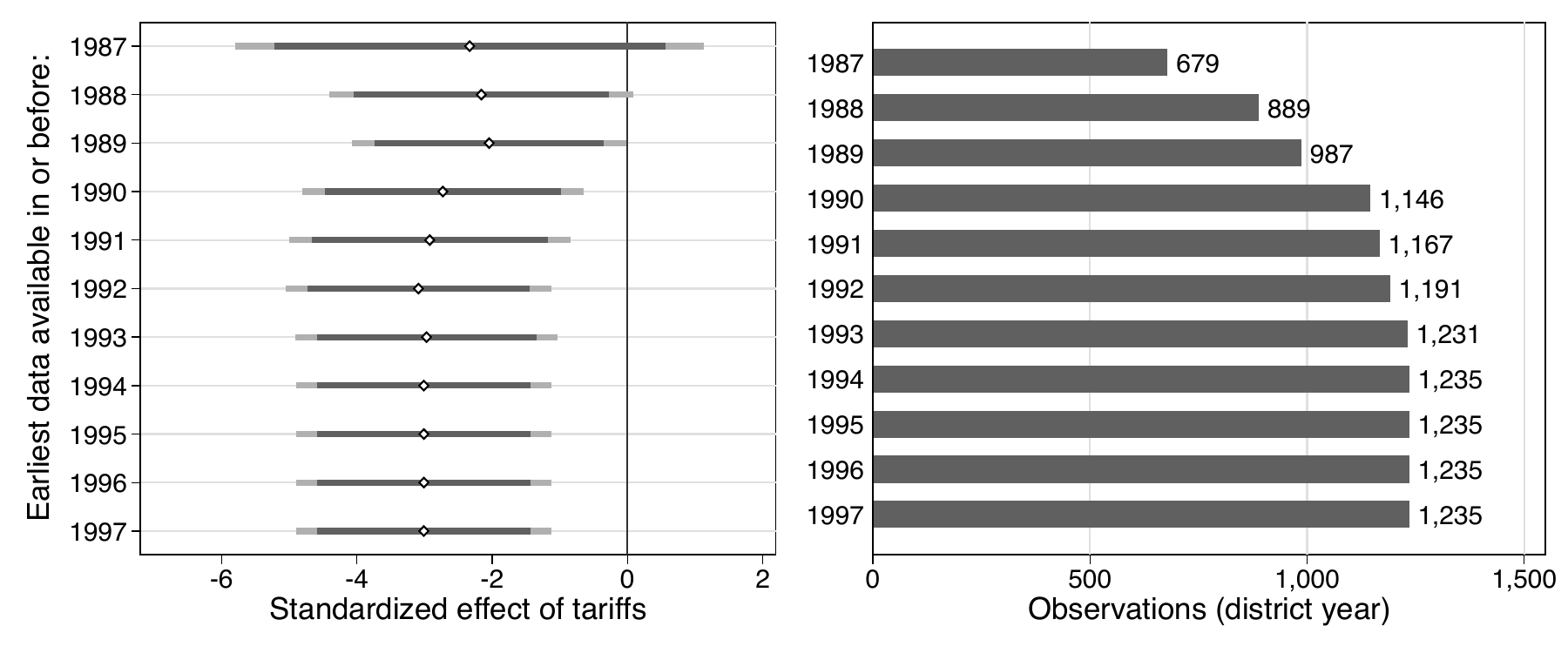}
    \caption{Applying sample restrictions at the district level}
    \label{fig:missingness_district}
  \end{subfigure}
  \begin{subfigure}{1\textwidth}
    \includegraphics[width=\linewidth]{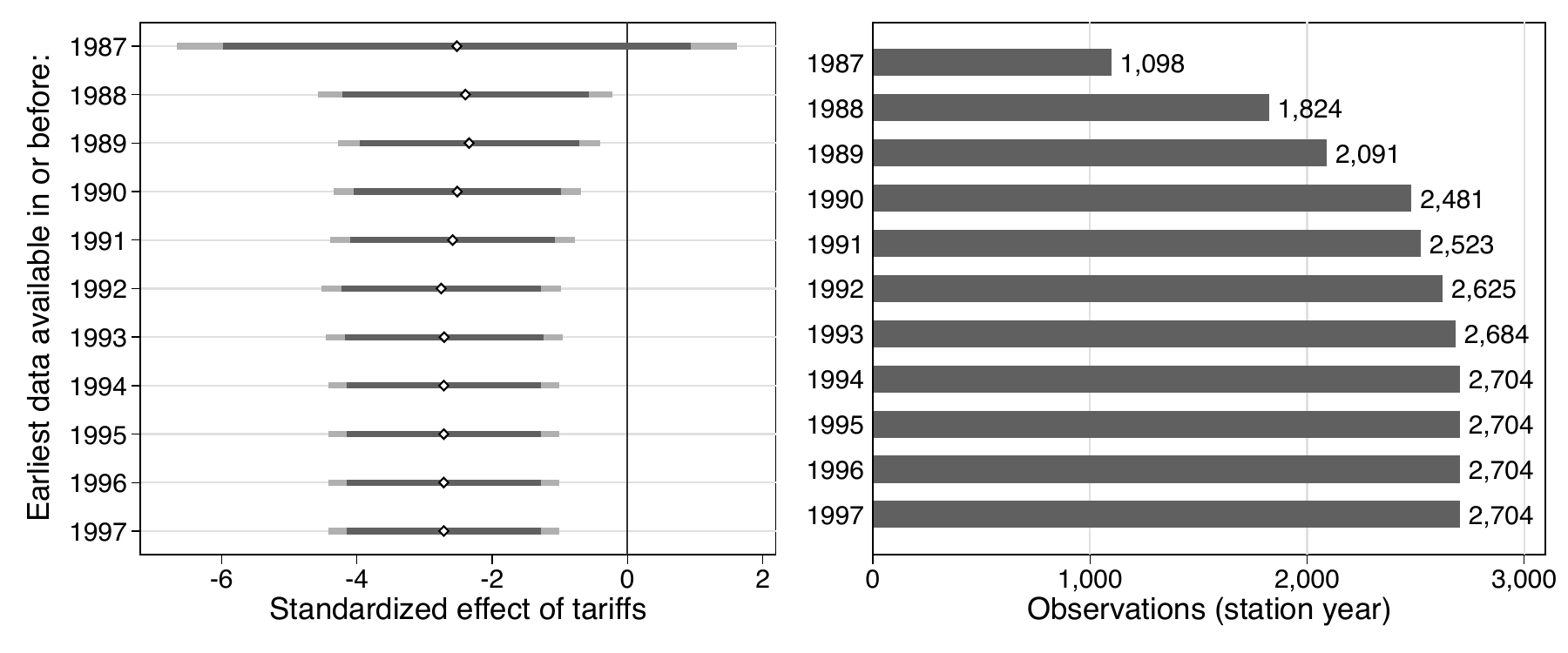}
    \caption{Applying sample restrictions at the station level}
    \label{fig:missingness_station}
  \end{subfigure}

  \caption{\raggedright Figure illustrates the effect of restricting the sample to districts (top panels) or monitoring stations (bottom panels) with the earliest pollution observations reported in or before a given year. The left-hand side figures show the estimated effect of tariffs on the pollution index when applying these sample restrictions. Point estimates are from regressions of the pollution index on the tariff exposure measure, district (or station) and year fixed effects, standard errors clustered by district. 90\% confidence intervals in dark gray; 95\% confidence intervals in lighter gray.  The right-hand side figures plot the number of observations. Our baseline estimates correspond to the row labeled ``1990'', where we restrict our sample to districts or monitoring stations with any pre-reform pollution data. The point estimates remain very stable with progressively stricter sample restrictions, becoming statistically insignificant only with the most stringent sample restrictions, which exclude more than half the station-level observations. \label{fig:missingness}}
  
\end{figure}


\begin{figure}[htbp!]
    \centering
    \includegraphics[width=1\textwidth]{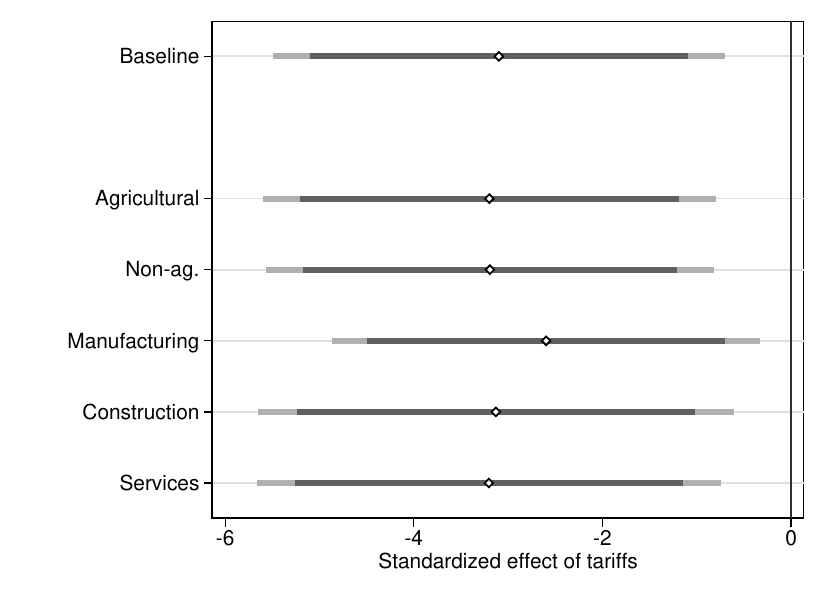}
    \caption{\raggedright The figure shows the estimated effect of tariffs on our index. Baseline shows the original effect using the model specified in Equation \ref{eqn:main_model}. The other estimates come from  models where year fixed effects are interacted with a district level variable from the NSS in 1987. Baseline regression includes district and year fixed effects, standard errors clustered by district. Remaining regressions include district fixed effects and interacted year fixed effects, standard errors clustered by district. 90\% confidence intervals in dark gray; 95\%  confidence intervals in lighter gray.}
    \label{fig:initial_nss}
\end{figure}

\begin{figure}[htbp!]
    \centering
    \includegraphics[width=1\textwidth]{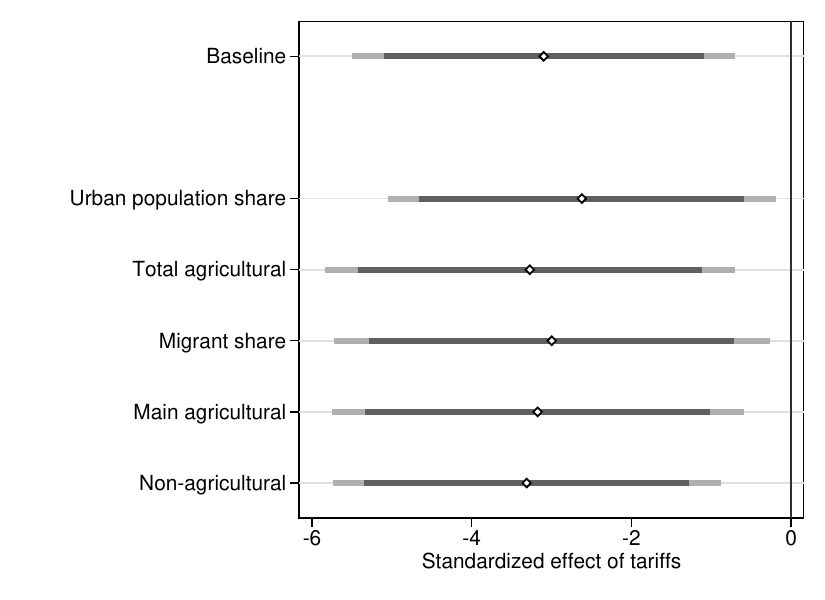}
    \caption{\raggedright The figure shows the estimated effect of tariffs on our index. Baseline shows the original effect using the model specified in Equation \ref{eqn:main_model}. The other estimates come from  models where year fixed effects are interacted with a district level variable from the Primary census abstract in 1991. Baseline regression includes district and year fixed effects, standard errors clustered by district. Remaining regressions include district fixed effects and interacted year fixed effects, standard errors clustered by district. 90\% confidence intervals in dark gray; 95\%  confidence intervals in lighter gray.}
    \label{fig:init_pca}
\end{figure}

\begin{figure}[htbp!]
    \centering
    \includegraphics[width=1\textwidth]{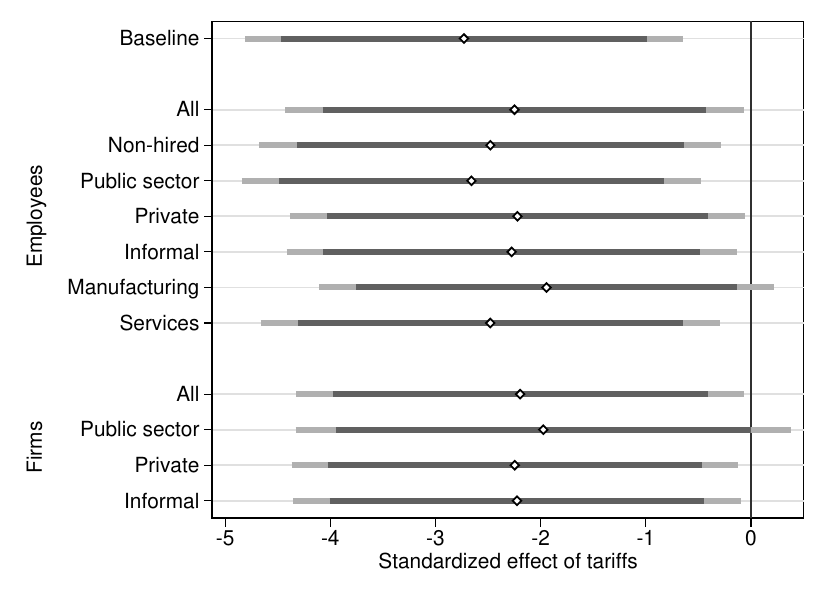}
    \caption{\raggedright The figure shows the estimated effect of tariffs on our index. Baseline shows the original effect using the model specified in Equation \ref{eqn:main_model}. The other estimates come from  models where year fixed effects are interacted with a district level variable from the Economic census in 1990. Baseline regression includes district and year fixed effects, standard errors clustered by district. Remaining regressions include district fixed effects and interacted year fixed effects, standard errors clustered by district. 90\% confidence intervals in dark gray; 95\%  confidence intervals in lighter gray.}
    \label{fig:init_ec}
\end{figure}

\clearpage


\begin{table}[p!]
    \centering
    \begin{threeparttable}
      \caption{Correlation between the water quality variables and index.}
      \input{tables/tableA1}
      \label{table:corr}
      \begin{tablenotes}
      \item \emph{Notes} Pairwise correlation coefficients for water pollution and quality metrics.
    \end{tablenotes}
    \end{threeparttable}
\end{table}

\begin{table}[p!]
    \centering
    \begin{threeparttable}
      \caption{Balance table}
      \input{tables/balance_table_early_ttest}
      \label{table:balance_early}
      \begin{tablenotes}
      \item \emph{Notes} Table compares sample districts to other districts in India. The main sample comprises districts with non-missing pre-reform water pollution data and tariff data. Difference signifies the raw difference in means between sample and excluded districts. The reported $p$ value tests the hypothesis of no difference between sample and excluded districts, assuming unequal variance (heteroskedasticity). Variables in Panels A and B are shares; variables in Panel C are reported in logs. 
    \end{tablenotes}
    \end{threeparttable}
\end{table}

\begin{sidewaystable}[p!]
    \centering
    \begin{threeparttable}
            \caption{Future tariff changes and pre-reform pollution levels}
            \input{tables/tableA2}
            \label{table:lvl_results}
    \begin{tablenotes}
      \small
      \item \emph{Notes} Robust standard errors in parentheses. Measures of water pollution are normalized to the pre-reform period. Tariff change is the difference between average pre-reform (before 1991) tariff exposure and average post-reform (1991 and after) tariff exposure. \mbox{** $p$ $<$ 0.05.}
    \end{tablenotes}
    \end{threeparttable}
\end{sidewaystable}

\begin{table}[p!]
    \centering
    \begin{threeparttable}
      \caption{Falsification exercises using data from \textcite{topalova2010factor}}
      \input{tables/tableA3_ours}
      \label{table:topalova_fals}
      \begin{tablenotes}
      \item \emph{Notes} Table uses data from  \textcite{topalova2010factor}. All regressions use the measure of tariff exposure we use throughout the paper, unit fixed effects, and a time dummy. Regressions use two pre-reform periods, with the tariff variable for the falsification test constructed by assigning post-reform tariffs to the second pre-reform period. Following \textcite{topalova2010factor}, standard errors (in parentheses) are clustered by state-year. Regressions are weighted by the number of households in a region (rural sample) or district (urban sample).
    \end{tablenotes}
    \end{threeparttable}
\end{table}

\begin{sidewaystable}[p!]
    \centering
    \begin{threeparttable}
      \caption{Comparing performance of traded tariff exposure and tariff exposure in falsification exercises using data from \textcite{topalova2010factor}}
      \input{tables/tableA3_v2}
      \label{table:topalova_fals_IV}
      \begin{tablenotes}
      \item \emph{Notes} Table uses data from \textcite{topalova2010factor}. Columns 1, 4, 7, and 10 replicate the falsification tests from Table \ref{table:topalova_fals} that correspond most closely to our analysis. Columns 2, 5, 8, and 11 add controls for initial conditions interacted with the post-reform dummy, as does \textcite{topalova2010factor}. Adding these controls tends to make the falsification tests less reassuring. Columns 3, 6, 9, and 12 use non-traded tariff exposure as an instrument for tariff exposure, retaining the controls for initial conditions, replicating the main specifications in \citeauthor{topalova2010factor} (\citeyear{topalova2010factor}, see Tables 3A and 3B, column 5). For urban areas, the falsification tests fail when using traded tariff exposure as an instrument. As in \textcite{topalova2010factor}, standard errors (in parentheses) are clustered by state-year. Regressions are weighted by the number of households in a region (rural sample) or district (urban sample). \mbox{* $p$ $<$ 0.10, ** $p$ $<$ 0.05}.
    \end{tablenotes}
    \end{threeparttable}
\end{sidewaystable}

\begin{table}[p!]
\centering
\begin{threeparttable}
    \caption{Using traded tariff exposure as an instrument for tariff exposure}
    \input{tables/tableA4}
    \label{table:iv_results}
    \begin{tablenotes}
      \small
      \item \emph{Notes} Column 1 shows main OLS specification for reference; column 2 shows first stage results; column 3 shows 2SLS results. The effective F-statistic follows \textcite{olea2013robust}. Standard errors (in parentheses) are clustered at the district level. Regressions include district and year fixed effects. All measures of water pollution are normalized to the pre-reform period. \mbox{*** $p$ $<$ 0.01.}
    \end{tablenotes}
\end{threeparttable}
\end{table}

\begin{sidewaystable}[p!]
    \centering
    \begin{threeparttable}
      \caption{Falsification exercise}
      \input{tables/table_fals_all}
      \label{table:fals_table_all}
    \begin{tablenotes}
      \small
      \item \emph{Notes} Results from the falsification exercise, where each water quality metric and the water pollution index are regressed on a four-year lead of the tariff measure. Sample includes data for years between 1987 and 1993. More extreme values suggest larger effects of trade on pollution. More extreme values suggest larger effects of trade on pollution. For DO, the scale is reversed, since DO negatively correlates with pollution levels. Standard errors (in parentheses) are clustered at the district level. All measures of water pollution are normalized to the pre-reform period. Year and district fixed effects are included in all regressions. \mbox{* $p$ $<$ 0.10}.
    \end{tablenotes}
    \end{threeparttable}
\end{sidewaystable}

\clearpage


\begin{sidewaystable}[p!]
    \centering
    \begin{threeparttable}
      \caption{Trade liberalization and population and employment outcomes -- Population census abstract}
      \input{tables/table_combined_pca}
      \label{table:pca}
      \begin{tablenotes}
      \item \emph{Notes} Table uses data from the Primary Census Abstract, provided and harmonized over time by \textcite{liu2023climate}. Outcomes are: log population counts (columns 1 and 2); share of the total population that is urban (column 3); share of the male population who are rural to urban within-district migrants (column 4); share of the labor force employed as agricultural laborers (column 5); share of the labor force employed as agricultural labors as their primary occupation (column 6); and share of the labor force in non-agricultural employment (column 7). Note that for results in column 5 and 7 the implicit omitted category is cultivators, who farm land at their own risk. District boundaries are harmonized following \textcite{liu2023climate}. Sample contains districts with non-missing census and tariff exposure data. Standard errors (in parentheses) are clustered at the district level. Regressions include district and year fixed effects. \mbox{* $p$ $<$ 0.10, ** $p$ $<$ 0.05, *** $p$ $<$ 0.01}.
    \end{tablenotes}
    \end{threeparttable}
\end{sidewaystable}

\begin{sidewaystable}[p!]
    \centering
    \begin{threeparttable}
       \caption{Trade liberalization and employment -- Economic census}
      \input{tables/table_combined_emp}
      \label{table:ec_all}
      \begin{tablenotes}
      \item \emph{Notes} Table uses data from the economic census \parencite{ecindia}, provided and harmonized over time by \textcite{asher2021development}. Outcomes are log employee counts for: all sectors (column 1), the self-employed (column 2), government-owned firms (column 3), private firms (column 4), the informal sector (column 5), the manufacturing sector (column 6), and the service sector (column 7). Sample contains districts with non-missing census and tariff exposure data. Standard errors (in parentheses) are clustered at the district level. Year and district fixed effects are included in all regressions. 
    \end{tablenotes}
    \end{threeparttable}
\end{sidewaystable}

\begin{table}[p!]
    \centering
    \begin{threeparttable}
      \caption{Trade liberalization and the number of firms -- Economic census}

      \input{tables/table_combined_firm}
      \label{table:ec_firm_m}
      \begin{tablenotes}

  \item \emph{Notes} Table uses data from the economic census \parencite{ecindia}, provided and harmonized over time by \textcite{asher2021development}. Outcomes are log firm counts for: all sectors (column 1), government-owned firms (column 2), private firms (column 3), and informal sector firms (column 4). Sample contains districts with non-missing census data, water pollution data, and tariff exposure data. Standard errors (in parentheses) are clustered at the district level. Year and district fixed effects are included in all regressions. 
    \end{tablenotes}
    \end{threeparttable}
\end{table}

\begin{landscape}
\begin{table}[p!]
\centering
\begin{threeparttable}
    \caption{Non-tariff barriers, industrial licensing, and FDI openness}
    \input{tables/table_ntb}
    \label{table:ntb}
    \begin{tablenotes}
      \item \emph{Notes} Regressions include the listed variables and district and year fixed effects. Data on non-tariff barriers, FDI, and licensing are from \textcite{topalova2010factor}. The sample includes data from 1987 and 1997, and the subsample of districts for which \textcite{topalova2010factor} reports data on the additional variables. Standard errors (in parentheses) are clustered at the district level.  \mbox{* $p$ $<$ 0.10, ** $p$ $<$ 0.05, *** $p$ $<$ 0.01}.
    \end{tablenotes}
\end{threeparttable}
\end{table}
\end{landscape}

%% file: tables/tableA1.tex
\begin{tabular}{l*{9}{c}}
\toprule
                &\multicolumn{9}{c}{}                                                                     \\
                &      BOD& Chloride&       DO& Hardness&  Sulfate&      TDS&Tot. coli.&Turbidity&    Index\\
\midrule
BOD             &     1.00&         &         &         &         &         &         &         &         \\
Chloride        &     0.25&     1.00&         &         &         &         &         &         &         \\
DO              &     0.51&     0.27&     1.00&         &         &         &         &         &         \\
Hardness        &     0.32&     0.57&     0.29&     1.00&         &         &         &         &         \\
Sulfate         &     0.31&     0.73&     0.28&     0.56&     1.00&         &         &         &         \\
TDS             &     0.39&     0.81&     0.36&     0.64&     0.79&     1.00&         &         &         \\
Tot. coli.      &     0.22&     0.23&     0.18&     0.19&     0.22&     0.35&     1.00&         &         \\
Turbidity       &     0.12&     0.17&    -0.07&     0.07&     0.26&     0.11&     0.09&     1.00&         \\
Index           &     0.63&     0.55&     0.68&     0.62&     0.55&     0.51&     0.52&     0.47&     1.00\\
\bottomrule
\end{tabular}

%% file: tables/balance_table_early_ttest.tex
\begin{tabular}{lcccc}
\toprule
& Sample & Non-sample & Difference & $p$ value \\
\midrule \multicolumn{5}{c}{\textbf{Panel A: Primary Census Abstract (1991)}} \\ \addlinespace
Urban population share&        0.25&        0.21&        0.03&       0.080\\
Migrant share       &        0.03&        0.02&        0.01&       $<$0.001~~~~\\
Total agricultural  &        0.25&        0.21&        0.05&       0.001\\
Main agricultural   &        0.25&        0.20&        0.06&       $<$0.001~~~~\\
Non-agricultural    &        0.30&        0.30&        0.00&       0.941\\
\addlinespace Observations & 87 & 192 & & \\

\midrule \multicolumn{5}{c}{\textbf{Panel B: National Sample Survey - Employment (1987)}} \\ \addlinespace
Agricultural        &        0.54&        0.54&       0.00&       0.937\\
Non-ag.             &        0.39&        0.37&        0.01&       0.544\\
Manufacturing       &        0.11&        0.09&        0.02&       0.024\\
Construction        &        0.03&        0.05&       -0.02&       0.004\\
Services            &        0.23&        0.22&        0.00&       0.720\\
\addlinespace Observations & 55 & 232 & & \\

\midrule \multicolumn{5}{c}{\textbf{Panel C: Economic Census (1990)}} \\ \addlinespace
Employed - all      &       11.35&       11.23&        0.13&       0.406\\
Non-hired           &        9.95&        9.57&        0.39&       0.001\\
Public sector       &        9.45&        9.26&        0.19&       0.204\\
Private             &       11.17&       11.05&        0.12&       0.427\\
Informal            &       10.92&       10.87&        0.06&       0.701\\
Firms - all         &       10.45&       10.43&        0.02&       0.879\\
Govt                &        7.71&        7.71&       0.00&       0.989\\
Private             &       10.37&       10.34&        0.03&       0.861\\
Informal            &       10.40&       10.39&        0.01&       0.932\\
Manufacturing       &       10.23&       10.13&        0.09&       0.519\\
Services            &       10.89&       10.76&        0.13&       0.399\\
\addlinespace Observations & 113 & 154 & & \\
\bottomrule
\end{tabular}

%% file: tables/tableA2.tex
{
\def\sym#1{\ifmmode^{#1}\else\(^{#1}\)\fi}
\begin{tabular}{l*{9}{c}}
\toprule
                &\multicolumn{1}{c}{(1)}&\multicolumn{1}{c}{(2)}&\multicolumn{1}{c}{(3)}&\multicolumn{1}{c}{(4)}&\multicolumn{1}{c}{(5)}&\multicolumn{1}{c}{(6)}&\multicolumn{1}{c}{(7)}&\multicolumn{1}{c}{(8)}&\multicolumn{1}{c}{(9)}\\
                &\multicolumn{1}{c}{Index}&\multicolumn{1}{c}{BOD}&\multicolumn{1}{c}{Chloride}&\multicolumn{1}{c}{DO}&\multicolumn{1}{c}{Hardness}&\multicolumn{1}{c}{Sulfate}&\multicolumn{1}{c}{TDS}&\multicolumn{1}{c}{Tot. coliform}&\multicolumn{1}{c}{Turbidity}\\
\midrule
Tariff change -- 1987-1997&    -2.81         &    -7.60         &    -2.41         &   -11.16\sym{**} &    -1.45         &     2.75         &    -2.28         &    -2.20         &     5.61         \\
                &   (2.91)         &   (7.00)         &   (5.08)         &   (4.89)         &   (4.04)         &   (4.84)         &   (4.51)         &   (4.95)         &   (4.34)         \\
\midrule
N (observations)&      117         &      117         &      117         &      116         &      117         &      114         &      114         &      105         &      108         \\
N (districts)   &      117         &      117         &      117         &      116         &      117         &      114         &      114         &      105         &      108         \\
\bottomrule
\end{tabular}
}

%% file: tables/tableA3_ours.tex
{
\def\sym#1{\ifmmode^{#1}\else\(^{#1}\)\fi}
\begin{tabular}{l*{4}{c}}
\toprule
                &\multicolumn{2}{c}{Rural}                               &\multicolumn{2}{c}{Urban}\\
                \cmidrule(lr){2-3} \cmidrule(lr){4-5} 
                &\multicolumn{1}{c}{(1)}&\multicolumn{1}{c}{(2)}&\multicolumn{1}{c}{(3)}&\multicolumn{1}{c}{(4)}\\
                &\multicolumn{1}{c}{Poverty rate}&\multicolumn{1}{c}{Consumption}&\multicolumn{1}{c}{Poverty rate}&\multicolumn{1}{c}{Consumption}\\
\midrule
Tariff          &    0.031         &    0.165         &   -0.220         &   -0.065         \\
                &  (0.472)         &  (0.295)         &  (0.354)         &  (0.309)         \\
\midrule
Model           &      OLS         &      OLS         &      OLS         &      OLS         \\
N (observations)&      128         &      128         &      125         &      125         \\
\bottomrule
\end{tabular}
}

%% file: tables/tableA3_v2.tex
{
\def\sym#1{\ifmmode^{#1}\else\(^{#1}\)\fi}
\begin{tabular}{l*{12}{c}}
\toprule
                &\multicolumn{6}{c}{Rural} &\multicolumn{6}{c}{Urban} \\
\cmidrule(lr){2-7} \cmidrule(lr){8-13}  
                &\multicolumn{3}{c}{Consumption} &\multicolumn{3}{c}{Poverty rate}  
                &\multicolumn{3}{c}{Consumption} &\multicolumn{3}{c}{Poverty rate}  \\
\cmidrule(lr){2-4} \cmidrule(lr){5-7} \cmidrule(lr){8-10} \cmidrule(lr){11-13}
                &\multicolumn{1}{c}{(1)}&\multicolumn{1}{c}{(2)}&\multicolumn{1}{c}{(3)}
                &\multicolumn{1}{c}{(4)}&\multicolumn{1}{c}{(5)}&\multicolumn{1}{c}{(6)}
                &\multicolumn{1}{c}{(7)}&\multicolumn{1}{c}{(8)}&\multicolumn{1}{c}{(9)}
                &\multicolumn{1}{c}{(10)}&\multicolumn{1}{c}{(11)}&\multicolumn{1}{c}{(12)}\\
\midrule
Tariff          &    0.031         &   -0.447         &    0.038         &    0.165         &    0.040         &   -0.085         &   -0.220         &    2.022         &    4.478\sym{*}  &   -0.065         &   -2.297         &   -5.629\sym{**} \\
                &  (0.472)         &  (0.465)         &  (1.000)         &  (0.295)         &  (0.224)         &  (0.463)         &  (0.354)         &  (1.270)         &  (2.349)         &  (0.309)         &  (1.462)         &  (2.494)         \\
\midrule
Model           &      OLS         &      OLS         &     2SLS         &      OLS         &      OLS         &     2SLS         &      OLS         &      OLS         &     2SLS         &      OLS         &      OLS         &     2SLS         \\
Controls        &       No         &      Yes         &      Yes         &       No         &      Yes         &      Yes         &       No         &      Yes         &      Yes         &       No         &      Yes         &      Yes         \\
N (observations)&      128         &      128         &      128         &      128         &      128         &      128         &      125         &      125         &      125         &      125         &      125         &      125         \\
\bottomrule
\end{tabular}
}

%% file: tables/tableA4.tex
{
\def\sym#1{\ifmmode^{#1}\else\(^{#1}\)\fi}
\begin{tabular}{l*{3}{c}}
\toprule
                    &\multicolumn{1}{c}{(1)}&\multicolumn{1}{c}{(2)}&\multicolumn{1}{c}{(3)}\\
                    &\multicolumn{1}{c}{Index}&\multicolumn{1}{c}{Tariff}&\multicolumn{1}{c}{Index}\\
\midrule
Tariff              &       -2.73\sym{**} &                     &       -0.83         \\
                    &      (1.05)         &                     &      (2.19)         \\
\addlinespace
Traded tariff       &                     &        0.11\sym{***}&                     \\
                    &                     &      (0.01)         &                     \\
\midrule
N (observations)    &        1146         &        1146         &        1146         \\
N (districts)       &         117         &         117         &         117         \\
Model               &         OLS         &         OLS         &        2SLS         \\
Effective F-statistic&                     &       53.48         &                     \\
Chi-sq p-value      &                     &                     &        0.39         \\
Adjusted R-squared  &                     &        0.92         &                     \\
Within R-squared    &                     &        0.16         &                     \\
\bottomrule
\end{tabular}
}

%% file: tables/table_fals_all.tex
{
\def\sym#1{\ifmmode^{#1}\else\(^{#1}\)\fi}
\begin{tabular}{l*{9}{c}}
\toprule
                &\multicolumn{1}{c}{(1)}&\multicolumn{1}{c}{(2)}&\multicolumn{1}{c}{(3)}&\multicolumn{1}{c}{(4)}&\multicolumn{1}{c}{(5)}&\multicolumn{1}{c}{(6)}&\multicolumn{1}{c}{(7)}&\multicolumn{1}{c}{(8)}&\multicolumn{1}{c}{(9)}\\
                &\multicolumn{1}{c}{Index}&\multicolumn{1}{c}{BOD}&\multicolumn{1}{c}{Chloride}&\multicolumn{1}{c}{DO}&\multicolumn{1}{c}{Hardness}&\multicolumn{1}{c}{Sulfate}&\multicolumn{1}{c}{TDS}&\multicolumn{1}{c}{Tot. coliform}&\multicolumn{1}{c}{Turbidity}\\
\midrule
Tariff\_{t+4}    &     0.41         &    -0.44         &    -2.48         &    -0.46         &     3.00         &    -4.72         &    -5.41\sym{*}  &     1.91         &    -2.05         \\
                &   (1.54)         &   (2.44)         &   (3.04)         &   (3.07)         &   (2.25)         &   (4.47)         &   (3.18)         &   (1.98)         &   (2.10)         \\
\midrule
N (observations)&      678         &      663         &      677         &      673         &      678         &      650         &      528         &      579         &      628         \\
N (districts)   &      116         &      116         &      116         &      116         &      116         &      115         &      113         &      105         &      109         \\
\bottomrule
\end{tabular}
}

%% file: tables/table_combined_pca.tex
{
\def\sym#1{\ifmmode^{#1}\else\(^{#1}\)\fi}
\begin{tabular}{l*{7}{c}}
\toprule
                &\multicolumn{1}{c}{(1)}&\multicolumn{1}{c}{(2)}&\multicolumn{1}{c}{(3)}&\multicolumn{1}{c}{(4)}&\multicolumn{1}{c}{(5)}&\multicolumn{1}{c}{(6)}&\multicolumn{1}{c}{(7)}\\
                &\multicolumn{1}{c}{\shortstack{Urban\\population}}&\multicolumn{1}{c}{\shortstack{Total\\population}}&\multicolumn{1}{c}{\shortstack{Urban\\pop. share}}&\multicolumn{1}{c}{\shortstack{Migrant\\share}}&\multicolumn{1}{c}{\shortstack{Tot.\\agricultural}}&\multicolumn{1}{c}{\shortstack{Main\\agricultural}}&\multicolumn{1}{c}{Non-agricultural}\\
\midrule \multicolumn{7}{l}{\textbf{Panel A: Matched sample}} \\
Tariff          &    -0.44         &     0.29         &    -0.37\sym{**} &    -0.07         &     0.77\sym{***}&     0.08         &    -0.42\sym{**} \\
                &   (0.43)         &   (0.25)         &   (0.18)         &   (0.05)         &   (0.22)         &   (0.17)         &   (0.20)         \\
\midrule
Mean            &    13.45         &    14.96         &     0.26         &     0.03         &     0.28         &     0.21         &     0.34         \\
N (observations)&      174         &      174         &      174         &      174         &      174         &      174         &      174         \\
N (districts)   &       87         &       87         &       87         &       87         &       87         &       87         &       87         \\
\midrule

\multicolumn{7}{l}{\textbf{Panel B: All districts}} \\
Tariff          &     0.10         &     0.27\sym{***}&    -0.13\sym{**} &    -0.03\sym{*}  &     0.35\sym{**} &     0.03         &    -0.18         \\
                &   (0.16)         &   (0.07)         &   (0.05)         &   (0.02)         &   (0.14)         &   (0.09)         &   (0.16)         \\
\midrule
Mean            &    13.20         &    14.82         &     0.23         &     0.02         &     0.25         &     0.19         &     0.32         \\
N (observations)&      476         &      478         &      478         &      478         &      478         &      478         &      478         \\
N (districts)   &      238         &      239         &      239         &      239         &      239         &      239         &      239         \\
\bottomrule
\end{tabular}
}

%% file: tables/table_combined_emp.tex
{
\def\sym#1{\ifmmode^{#1}\else\(^{#1}\)\fi}
\begin{tabular}{l*{7}{c}}
\toprule
                &\multicolumn{1}{c}{(1)}&\multicolumn{1}{c}{(2)}&\multicolumn{1}{c}{(3)}&\multicolumn{1}{c}{(4)}&\multicolumn{1}{c}{(5)}&\multicolumn{1}{c}{(6)}&\multicolumn{1}{c}{(7)}\\
                &\multicolumn{1}{c}{All sectors}&\multicolumn{1}{c}{Non-hired}&\multicolumn{1}{c}{Public sector}&\multicolumn{1}{c}{Private}&\multicolumn{1}{c}{Informal}&\multicolumn{1}{c}{Manufacturing}&\multicolumn{1}{c}{Services}\\
\midrule \multicolumn{7}{l}{\textbf{Panel A: Matched sample}} \\
Tariff          &    -0.16         &    -0.45         &     1.51         &    -0.26         &    -1.90         &     0.32         &    -0.63         \\
                &   (3.59)         &   (2.77)         &   (3.66)         &   (3.61)         &   (4.05)         &   (3.97)         &   (3.48)         \\
\midrule
Mean            &    11.72         &    10.70         &     9.69         &    11.56         &    11.39         &    10.51         &    11.30         \\
N (observations)&      226         &      222         &      226         &      226         &      194         &      226         &      226         \\
N (districts)   &      113         &      111         &      113         &      113         &       97         &      113         &      113         \\
\midrule

\multicolumn{7}{l}{\textbf{Panel B: All districts}} \\
Tariff          &    -0.20         &     0.81         &    -0.21         &    -0.14         &    -0.84         &    -0.76         &    -0.11         \\
                &   (0.65)         &   (0.63)         &   (0.73)         &   (0.66)         &   (0.59)         &   (0.66)         &   (0.63)         \\
\midrule
Mean            &    11.56         &    10.48         &     9.50         &    11.39         &    11.27         &    10.35         &    11.13         \\
N (observations)&      536         &      530         &      536         &      536         &      474         &      534         &      536         \\
N (districts)   &      268         &      265         &      268         &      268         &      237         &      267         &      268         \\
\bottomrule
\end{tabular}
}

%% file: tables/table_combined_firm.tex
{
\def\sym#1{\ifmmode^{#1}\else\(^{#1}\)\fi}
\begin{tabular}{l*{4}{c}}
\toprule
                &\multicolumn{1}{c}{(1)}&\multicolumn{1}{c}{(2)}&\multicolumn{1}{c}{(3)}&\multicolumn{1}{c}{(4)}\\
                &\multicolumn{1}{c}{All sectors}&\multicolumn{1}{c}{Government}&\multicolumn{1}{c}{Private}&\multicolumn{1}{c}{Informal}\\
\midrule \multicolumn{5}{l}{\textbf{Panel A: Matched sample}} \\
Tariff          &    -0.12         &     3.13         &    -0.38         &    -0.43         \\
                &   (3.38)         &   (3.84)         &   (3.39)         &   (3.72)         \\
\midrule
Mean            &    10.82         &     7.90         &    10.75         &    10.73         \\
N (observations)&      226         &      226         &      226         &      194         \\
N (districts)   &      113         &      113         &      113         &       97         \\
\midrule

\multicolumn{5}{l}{\textbf{Panel B: All districts}} \\
Tariff          &     0.10         &     0.17         &     0.10         &    -0.21         \\
                &   (0.62)         &   (0.86)         &   (0.64)         &   (0.48)         \\
\midrule
Mean            &    10.71         &     7.82         &    10.64         &    10.63         \\
N (observations)&      536         &      536         &      536         &      474         \\
N (districts)   &      268         &      268         &      268         &      237         \\
\bottomrule
\end{tabular}
}

%% file: tables/table_ntb.tex
{
\def\sym#1{\ifmmode^{#1}\else\(^{#1}\)\fi}
\begin{tabular}{l*{9}{c}}
\toprule
                &\multicolumn{1}{c}{(1)}&\multicolumn{1}{c}{(2)}&\multicolumn{1}{c}{(3)}&\multicolumn{1}{c}{(4)}&\multicolumn{1}{c}{(5)}&\multicolumn{1}{c}{(6)}&\multicolumn{1}{c}{(7)}&\multicolumn{1}{c}{(8)}&\multicolumn{1}{c}{(9)}\\
                &\multicolumn{1}{c}{Index}&\multicolumn{1}{c}{Index}&\multicolumn{1}{c}{Index}&\multicolumn{1}{c}{Index}&\multicolumn{1}{c}{Index}&\multicolumn{1}{c}{Index}&\multicolumn{1}{c}{Index}&\multicolumn{1}{c}{Index}&\multicolumn{1}{c}{Index}\\
\midrule
Tariff          &    -3.48\sym{***}&                  &                  &                  &                  &    -4.33\sym{**} &    -3.68\sym{***}&    -3.43\sym{***}&    -3.96\sym{**} \\
                &   (1.25)         &                  &                  &                  &                  &   (1.74)         &   (1.26)         &   (1.26)         &   (1.73)         \\
\addlinespace
NTBs            &                  &     3.18         &                  &                  &     3.12         &    -1.86         &                  &                  &    -1.47         \\
                &                  &   (2.12)         &                  &                  &   (2.26)         &   (2.66)         &                  &                  &   (2.73)         \\
\addlinespace
Unrestricted FDI&                  &                  &     0.04         &                  &     0.21         &                  &    -0.25         &                  &     0.17         \\
                &                  &                  &   (0.49)         &                  &   (0.56)         &                  &   (0.45)         &                  &   (0.50)         \\
\addlinespace
Industrial licensing&                  &                  &                  &     0.54\sym{**} &     0.67\sym{*}  &                  &                  &     0.52\sym{*}  &     0.57\sym{*}  \\
                &                  &                  &                  &   (0.26)         &   (0.35)         &                  &                  &   (0.27)         &   (0.33)         \\
\midrule
N (observations)&      134         &      134         &      134         &      134         &      134         &      134         &      134         &      134         &      134         \\
N (districts)   &       67         &       67         &       67         &       67         &       67         &       67         &       67         &       67         &       67         \\
\bottomrule
\end{tabular}
}

%% file: tradepollution.bib
@unpublished{fouquin2016back,
  title={Back to the future: international trade costs and the two globalizations},
  author={Fouquin, Michel and Hugot, Jules},
  year={2016},
  note={Centre d'etudes prospectives et d'informations internationales (CEPII) Working Paper N°2016-13}
}

@misc{india_nrcp_lok_sabha_2018,
  author       = {{Ministry of Environment, Forest and Climate Change, Government of India}},
  title        = {{Lok Sabha Unstarred Question No. 3970: ``National River Conservation Programme''}},
  year         = {2018},
  month        = aug,
  note         = {Answered on 10 August 2018 by the Ministry of Environment, Forest and Climate Change},
  publisher    = {Government of India},  
  howpublished = {\url{https://sansad.in}},
}

@article{rambachan2023more,
  title={A more credible approach to parallel trends},
  author={Rambachan, Ashesh and Roth, Jonathan},
  journal={Review of Economic Studies},
  volume={90},
  number={5},
  pages={2555--2591},
  year={2023}
}

@misc{MLInfomap2013,
  author       = {{ML Infomap}},
  title        = {{District Boundaries of India, 1961}},
  year         = {2013},
  publisher    = {ML Infomap},
  note         = {Dataset}
}

@article{lehner2013global,
  title={Global river hydrography and network routing: baseline data and new approaches to study the world's large river systems},
  author={Lehner, Bernhard and Grill, G{\"u}nther},
  journal={Hydrological Processes},
  volume={27},
  number={15},
  pages={2171--2186},
  year={2013},
  publisher={Wiley Online Library}
}

@article{olea2013robust,
  title={A robust test for weak instruments},
  author={Olea, Jos{\'e} Luis Montiel and Pflueger, Carolin},
  journal={Journal of Business \& Economic Statistics},
  volume={31},
  number={3},
  pages={358--369},
  year={2013},
  publisher={Taylor \& Francis}
}

@misc{IPUMS_India_1987_GIS,
  author       = {Steven Ruggles and Lara L. Cleveland and Rodrigo Lovat{\'o}n D{\'a}vila and Sula Sarkar and Matthew Sobek and Derek Burk and Dan E. Ehrlich and Jane Lee and Nate Merrill},
  title        = {{IPUMS International: Version 7.3 [dataset]. India - Year-Specific Second-Level GIS Boundary Files, 1987}},
  year         = {2023},
  publisher    = {IPUMS},
  address      = {Minneapolis, MN},
  url          = {https://international.ipums.org/international/gis_yrspecific_2nd.shtml},
  doi          = {10.18128/D020.V7.7},
  note         = {Accessed: 2025. Source: Ministry of Statistics and Programme Implementation (MOSPI)}
}

@article{Cook1977,
  author    = {Cook, R. Dennis},
  title     = {Detection of Influential Observations in Linear Regression},
  journal   = {Technometrics},
  year      = {1977},
  volume    = {19},
  number    = {1},
  pages     = {15--18}
}

@article{aghion2008unequal,
  title={{The unequal effects of liberalization: Evidence from dismantling the License Raj in India}},
  author={Aghion, Philippe and Burgess, Robin and Redding, Stephen J and Zilibotti, Fabrizio},
  journal={American Economic Review},
  volume={98},
  number={4},
  pages={1397--1412},
  year={2008},
  publisher={American Economic Association}
}

@article{choi2025cleanup,
  title={The cleanup of US manufacturing through pollution offshoring},
  author={Choi, Jaerim and Hyun, Jay and Kim, Gueyon and Park, Ziho},
  journal={Journal of International Economics},
  volume={154},
  pages={104046},
  year={2025}
}

@article{levinson2023developed,
  title={Are developed countries outsourcing pollution?},
  author={Levinson, Arik},
  journal={Journal of Economic Perspectives},
  volume={37},
  number={3},
  pages={87--110},
  year={2023}
}

@article{he2020watering,
  title={Watering down environmental regulation in {China}},
  author={He, Guojun and Wang, Shaoda and Zhang, Bing},
  journal={The Quarterly Journal of Economics},
  volume={135},
  number={4},
  pages={2135--2185},
  year={2020},
}

@book{excell2017thirsting,
  title={{Thirsting for justice: transparency and poor people's struggle for clean water in Indonesia, Mongolia, and Thailand.}},
  author={Excell, Carolee and Moses, Elizabeth},
  year={2017},
  publisher={World Resources Institute}
}

@article{lamers2013sulfide,
  title={Sulfide as a soil phytotoxin—a review},
  author={Lamers, Leon PM and Govers, Laura L and Janssen, Inge CJM and Geurts, Jeroen JM and Van der Welle, Marlies EW and Van Katwijk, Marieke M and Van der Heide, Tjisse and Roelofs, Jan GM and Smolders, Alfons JP},
  journal={Frontiers in Plant Science},
  volume={4},
  pages={268},
  year={2013},
  publisher={Frontiers Media SA}
}

@techreport{WHO2004_sulfate,
  author       = {{World Health Organization}},
  title        = {{Sulfate in Drinking-water: Background Document for Development of {WHO} Guidelines for Drinking-water Quality}},
  institution  = {World Health Organization},
  type         = {Background document},
  address      = {Geneva, Switzerland},
  year         = {2004},
  number       = {WHO/SDE/WSH/03.04/114}%,
  %note         = {Prepared during development of the 3rd edition of the WHO Guidelines for Drinking-water Quality}
}

@misc{usgs_chloride_salinity_dissolved2019,
  author       = {{United States Geological Survey}},
  title        = {Chloride, Salinity, and Dissolved Solids},
  howpublished = {\url{https://www.usgs.gov/mission-areas/water-resources/science/chloride-salinity-and-dissolved-solids}},
  year         = {2019},
 % month        = mar,
 % day          = {1},
 % note         = {Accessed: 2025-06-19}
}

@techreport{WHO2011hardness,
  title        = {{Hardness in Drinking-water: Background Document for Development of WHO Guidelines for Drinking-water Quality}},
  author       = {{World Health Organization}},
  institution  = {World Health Organization},
  year         = {2011},
  type         = {WHO/SDE/WSH/03.04/06},
  url          = {https://www.who.int/water_sanitation_health/dwq/chemicals/hardness.pdf},
}

@incollection{chapman2013surface,
  author    = {Chapman, P. J. and Kay, P. and Mitchell, G. N. and  Pitts, C.},
  title     = {Surface water quality},
  booktitle = {Water Resources: An Integrated Approach},
  editor    = {Holden, J.},
  publisher = {Routledge},
  year      = {2013},
  pages     = {79--122},
 % isbn      = {0415602823}
}

@book{chapman1996water,
  title={Water quality assessments: a guide to the use of biota, sediments and water in environmental monitoring},
  author={Chapman, Deborah V},
  year={1996},
  publisher={CRC Press}
}

@article{taylor2011buffalo,
  title={Buffalo hunt: International trade and the virtual extinction of the North American bison},
  author={Taylor, M Scott},
  journal={American Economic Review},
  volume={101},
  number={7},
  pages={3162--3195},
  year={2011},
}

@article{asher2021development,
  title={{Development research at high geographic resolution: an analysis of night-lights, firms, and poverty in India using the SHRUG open data platform}},
  author={Asher, Sam and Lunt, Tobias and Matsuura, Ryu and Novosad, Paul},
  journal={The World Bank Economic Review},
  volume={35},
  number={4},
  pages={845--871},
  year={2021},
  publisher={Oxford University Press}
}

@article{zak2021sulphate,
  title={Sulphate in freshwater ecosystems: A review of sources, biogeochemical cycles, ecotoxicological effects and bioremediation},
  author={Zak, Dominik and Hupfer, Michael and Cabezas, Alvaro and Jurasinski, Gerald and Audet, Joachim and Kleeberg, Andreas and McInnes, Robert and Kristiansen, S{\o}ren Munch and Petersen, Rasmus Jes and Liu, Haojie and others},
  journal={Earth-Science Reviews},
  volume={212},
  pages={103446},
  year={2021},
  publisher={Elsevier}
}

@incollection{world2003tds,
  title={Total dissolved solids in drinking-water: background document for development of WHO Guidelines for drinking-water quality},
  author={{World Health Organization}},
  booktitle={Total dissolved solids in drinking-water: background document for development of WHO Guidelines for drinking-water quality},
  publisher={World Health Organization},
  year={2003}
}

@techreport{granato2015methods,
  title={{Methods for evaluating potential sources of chloride in surface waters and groundwaters of the conterminous United States}},
  author={Granato, Gregory E and DeSimone, Leslie A and Barbaro, Jeffrey R and Jeznach, Lillian C},
  year={2015},
  institution={US Geological Survey}
}

@misc{ecindia,
  title={{Economic Census of India}},
  author={Central Statistics Organisation, Ministry of Statistics and Programme Implementation, Government of India},
  year={2013},
  note={{Accessed via SHRUG \url{https://www.devdatalab.org/shrug_download/}, v2.1.pakora}}
}

@article{liu2023climate,
  title={{Climate change and labor reallocation: Evidence from six decades of the Indian Census}},
  author={Liu, Maggie and Shamdasani, Yogita and Taraz, Vis},
  journal={American Economic Journal: Economic Policy},
  volume={15},
  number={2},
  pages={395--423},
  year={2023},
  publisher={American Economic Association 2014 Broadway, Suite 305, Nashville, TN 37203-2425}
}

@incollection{alsina2017effect,
  title={The effect of carbon taxes on emissions and carbon leakage: evidence from the {European Union}},
  author={Alsina-Pujols, Maria},
  booktitle={{The Green Market Transition}},
  pages={30--46},
  year={2017},
  publisher={Edward Elgar Publishing}
}

@article{aichele2015kyoto,
  title={Kyoto and carbon leakage: An empirical analysis of the carbon content of bilateral trade},
  author={Aichele, Rahel and Felbermayr, Gabriel},
  journal={Review of Economics and Statistics},
  volume={97},
  number={1},
  pages={104--115},
  year={2015},
  publisher={The MIT Press}
}

@article{grubb2022carbon,
  title={Carbon leakage, consumption, and trade},
  author={Grubb, Michael and Jordan, Nino David and Hertwich, Edgar and Neuhoff, Karsten and Das, Kasturi and Bandyopadhyay, Kaushik Ranjan and Van Asselt, Harro and Sato, Misato and Wang, Ranran and Pizer, William A and Oh, Hyungna},
  journal={Annual Review of Environment and Resources},
  volume={47},
  number={1},
  pages={753--795},
  year={2022}
}

@article{copeland1994north,
 author = {Brian R. Copeland and M. Scott Taylor},
 journal = {Quarterly Journal of Economics},
 number = {3},
 pages = {755--787},
 title = {{North-South Trade and the Environment}},
 volume = {109},
 year = {1994}
}

@article{copeland2008pollution,
  title={The pollution haven hypothesis},
  author={Copeland, Brian R},
  journal={Handbook on Trade and the Environment},
  volume={2},
  number={7},
  year={2008},
  publisher={Edward Elgar Publishing: Cheltenham, UK}
}

@article{Lepault2024,
    author = {Lepault, Claire},
    title = {Is urban wastewater treatment effective in {I}ndia? {E}vidence from water quality and infant mortality},
    year = {2024}
 }

@article{grossman1995economic,
  title={Economic growth and the environment},
  author={Grossman, Gene M and Krueger, Alan B},
  journal={Quarterly Journal of Economics},
  volume={110},
  number={2},
  pages={353--377},
  year={1995}
}

@article{anderson2008multiple,
  title={{Multiple inference and gender differences in the effects of early intervention: A reevaluation of the Abecedarian, Perry Preschool, and Early Training Projects}},
  author={Anderson, Michael L},
  journal={Journal of the American Statistical Association},
  volume={103},
  number={484},
  pages={1481--1495},
  year={2008}
}

@article{hasan2012trade,
  title={{Trade liberalization and unemployment: Theory and evidence from India}},
  author={Hasan, Rana and Mitra, Devashish and Ranjan, Priya and Ahsan, Reshad N},
  journal={Journal of Development Economics},
  volume={97},
  number={2},
  pages={269--280},
  year={2012}
}

@article{abman2020does,
  title={{Does free trade increase deforestation? The effects of regional trade agreements}},
  author={Abman, Ryan and Lundberg, Clark},
  journal={Journal of the Association of Environmental and Resource Economists},
  volume={7},
  number={1},
  pages={35--72},
  year={2020}
}

@online{samii,
    author = {Samii, Cyrus},
    title = {make\_index},
    publisher = {GitHub},
    journal = {GitHub repository},
    url = {https://github.com/cdsamii/make_index},
    year={2018}
}

@article{shapiro2018pollution,
  title={Why is pollution from {US} manufacturing declining? {T}he roles of environmental regulation, productivity, and trade},
  author={Shapiro, Joseph S and Walker, Reed},
  journal={American Economic Review},
  volume={108},
  number={12},
  pages={3814--3854},
  year={2018}
}

@article{he2016surface,
  title={{Surface water quality and infant mortality in China}},
  author={He, Guojun and Perloff, Jeffrey M},
  journal={Economic Development and Cultural Change},
  volume={65},
  number={1},
  pages={119--139},
  year={2016}
}

@article{goldberg2010multiproduct,
  title={{Multiproduct firms and product turnover in the developing world: Evidence from India}},
  author={Goldberg, Pinelopi K and Khandelwal, Amit K and Pavcnik, Nina and Topalova, Petia},
  journal={Review of Economics and Statistics},
  volume={92},
  number={4},
  pages={1042--1049},
  year={2010}
}

@article{edmonds2010trade,
  title={{Trade adjustment and human capital investments: Evidence from Indian tariff reform}},
  author={Edmonds, Eric V and Pavcnik, Nina and Topalova, Petia},
  journal={American Economic Journal: Applied Economics},
  volume={2},
  number={4},
  pages={42--75},
  year={2010}
}

@article{edmonds2009child,
  title={{Child labor and schooling in a globalizing world: Some evidence from urban India}},
  author={Edmonds, Eric V and Topalova, Petia and Pavcnik, Nina},
  journal={Journal of the European Economic Association},
  volume={7},
  number={2-3},
  pages={498--507},
  year={2009}
}

@article{sekhri2022agricultural,
  title={{Agricultural trade and depletion of groundwater}},
  author={Sekhri, Sheetal},
  journal={Journal of Development Economics},
  volume={156},
  pages={102800},
  year={2022}
}

@article{ebenstein2012consequences,
  title={{The consequences of industrialization: evidence from water pollution and digestive cancers in China}},
  author={Ebenstein, Avraham},
  journal={Review of Economics and Statistics},
  volume={94},
  number={1},
  pages={186--201},
  year={2012},
  publisher={The MIT Press}
}

@misc{grossman1991environmental,
  title={{Environmental impacts of a North American free trade agreement}},
  author={Grossman, Gene M and Krueger, Alan B},
  year={1991},
  publisher={National Bureau of Economic Research Cambridge, Mass., USA}
}

@article{keiser2019consequences,
  title={{Consequences of the Clean Water Act and the demand for water quality}},
  author={Keiser, David A and Shapiro, Joseph S},
  journal={Quarterly Journal of Economics},
  volume={134},
  number={1},
  pages={349--396},
  year={2019}
}

@article{cherniwchan2017tradeb,
  title={Trade and the environment: New methods, measurements, and results},
  author={Cherniwchan, Jevan and Copeland, Brian R and Taylor, M Scott},
  journal={Annual Review of Economics},
  volume={9},
  pages={59--85},
  year={2017}
}

@article{bombardini2020trade,
  title={{Trade, pollution and mortality in China}},
  author={Bombardini, Matilde and Li, Bingjing},
  journal={Journal of International Economics},
  volume={125},
  pages={103321},
  year={2020}
}

@article{cerra2002caused,
  title={{What caused the 1991 currency crisis in India?}},
  author={Cerra, Valerie and Saxena, Sweta Chaman},
  journal={IMF Staff Papers},
  volume={49},
  number={3},
  pages={395--425},
  year={2002}
}

@article{chen2020wto,
  title={{WTO accession, trade expansion, and air pollution: Evidence from China’s county-level panel data}},
  author={Chen, Shuai and Lin, Faqin and Yao, Xi and Zhang, Peng},
  journal={Review of International Economics},
  volume={28},
  number={4},
  pages={1020--1045},
  year={2020}
}

@article{cherniwchan2017trade,
  title={{Trade liberalization and the environment: Evidence from NAFTA and US manufacturing}},
  author={Cherniwchan, Jevan},
  journal={Journal of International Economics},
  volume={105},
  pages={130--149},
  year={2017}
}

@article{do2018can,
  title={{Can environmental policy reduce infant mortality? Evidence from the Ganga Pollution Cases}},
  author={Do, Quy-Toan and Joshi, Shareen and Stolper, Samuel},
  journal={Journal of Development Economics},
  volume={133},
  pages={306--325},
  year={2018}
}

@article{borusyak2022quasi,
  title={{Quasi-experimental shift-share research designs}},
  author={Borusyak, Kirill and Hull, Peter and Jaravel, Xavier},
  journal={Review of Economic Studies},
  volume={89},
  number={1},
  pages={181--213},
  year={2022}
}

@article{garg2018not,
  title={{(Not so) gently down the stream: River pollution and health in Indonesia}},
  author={Garg, Teevrat and Hamilton, Stuart E and Hochard, Jacob P and Kresch, Evan Plous and Talbot, John},
  journal={Journal of Environmental Economics and Management},
  volume={92},
  pages={35--53},
  year={2018}
}

@article{topalova2011trade,
  title={{Trade liberalization and firm productivity: The case of India}},
  author={Topalova, Petia and Khandelwal, Amit},
  journal={Review of Economics and Statistics},
  volume={93},
  number={3},
  pages={995--1009},
  year={2011}
}

@article{greenstone2014environmental,
  title={{Environmental regulations, air and water pollution, and infant mortality in India}},
  author={Greenstone, Michael and Hanna, Rema},
  journal={American Economic Review},
  volume={104},
  number={10},
  pages={3038--72},
  year={2014}
}

@inbook{khanna1995trade,
  title={{Trade Policy Reforms: The India Experience}},
  author={Khanna, Tejendra},
  editor={Naheed Kirmani and Chorng-Huey Wong},
 booktitle={{Trade policy issues: papers presented at the Seminar on Trade Policy Issues, FMI, Washington}},
chapter={5},
  pages={49--57},
  year={1995}
}

@book{chopra1995india,
  title={India: economic reform and growth},
  author={Chopra, Mr Ajai and Collyns, Mr Charles and Hemming, Mr Richard and Parker, Ms Karen Elizabeth and Chu, Woosik and Fratzscher, Mr Oliver},
  year={1995},
  publisher={International Monetary Fund}
}

@article{conley1999gmm,
  title={{GMM estimation with cross sectional dependence}},
  author={Conley, Timothy G},
  journal={Journal of Econometrics},
  volume={92},
  number={1},
  pages={1--45},
  year={1999}
}

@article{acharya2002india,
  title={{India: Crisis, Reforms and Growth in the Nineties}},
  author={Acharya, Shankar},
  year={2002}
}

@article{topalova2010factor,
  title={{Factor immobility and regional impacts of trade liberalization: Evidence on poverty from India}},
  author={Topalova, Petia},
  journal={American Economic Journal: Applied Economics},
  year={2010},
  volume={2},
  number={4},
  pages={1--41}
}

@book{worldbank2013india,
  title={India:
Diagnostic Assessment of Select Environmental Challenges. 
{A}n Analysis of Physical and Monetary Losses of
Environmental Health and Natural Resources. {Report No. 70004-IN.}},
  author={{World Bank}},
  year={2013},
  publisher={World Bank Publications}
}

@book{topalova2007,
  title={{7 Trade Liberalization, Poverty, and Inequality}},
  author={Topalova, Petia},
  year={2007},
  publisher={University of Chicago Press}
}

@article{kumar2017creating,
  title={{Creating Long Panels Using Census Data (1961-2001)}},
  author={Kumar, Hemanshu and Somanathan, Rohini},
  journal={Economic and Political Weekly},
  volume={52},
  number={29},
  pages={105--109},
  year={2017}
}

@article{martin2011energy,
  title={{Energy efficiency gains from trade: greenhouse gas emissions and India’s manufacturing sector}},
  author={Martin, Leslie A},
  journal={Mimeograph, Berkeley ARE},
  year={2011}
}

@article{sargaonkar2003development,
  title={{Development of an overall index of pollution for surface water based on a general classification scheme in Indian context.}},
  author={Sargaonkar, Aabha and Deshpande, Vijaya},
  journal={Environmental Monitoring \& Assessment},
  volume={89},
  number={1},
  year={2003}
}

@misc{goldberg2004trade,
  title={{Trade, inequality, and poverty: What do we know? Evidence from recent trade liberalization episodes in developing countries}},
  author={Goldberg, Pinelopi K and Pavcnik, Nina},
  year={2004},
  publisher={National Bureau of Economic Research Cambridge, Mass., USA}
}

@article{goldberg2010imported,
  title={{Imported intermediate inputs and domestic product growth: Evidence from India}},
  author={Goldberg, Pinelopi Koujianou and Khandelwal, Amit Kumar and Pavcnik, Nina and Topalova, Petia},
  journal={Quarterly Journal of Economics},
  volume={125},
  number={4},
  pages={1727--1767},
  year={2010}
}

@book{singh2017trade,
  title={{Trade Policy Reform in India since 1991}},
  author={Singh, Harsha Vardhana},
  year={2017},
  publisher={Brookings India Working Paper 02}
}

@article{copeland2004trade,
  title={{Trade, growth, and the environment}},
  author={Copeland, Brian R and Taylor, M Scott},
  journal={Journal of Economic Literature},
  volume={42},
  number={1},
  pages={7--71},
  year={2004}
}

@book{carr2008water,
  title={Water quality for ecosystem and human health},
  author={Carr, Genevi{\`e}ve M and Neary, James P},
  year={2008},
  publisher={UNEP/Earthprint}
}

@article{khan2008health,
  title={{Health risks of heavy metals in contaminated soils and food crops irrigated with wastewater in Beijing, China}},
  author={Khan, S and Cao, Q and Zheng, YM and Huang, YZ and Zhu, YG},
  journal={Environmental Pollution},
  volume={152},
  number={3},
  pages={686--692},
  year={2008}
}

@article{sharma2006heavy,
  title={{Heavy metal contamination in vegetables grown in wastewater irrigated areas of Varanasi, India.}},
  author={Sharma, RK and Agrawal, M and Marshall, F},
  journal={Bulletin of Environmental Contamination \& Toxicology},
  volume={77},
  number={2},
  year={2006}
}

@article{shapiro2016trade,
  title={Trade costs, {CO2}, and the environment},
  author={Shapiro, Joseph S},
  journal={American Economic Journal: Economic Policy},
  volume={8},
  number={4},
  pages={220--254},
  year={2016}
}

@article{chen2023does,
  title={Does export liberalization cause the agglomeration of pollution? {E}vidence from {C}hina},
  author={Chen, Xiaoping and Shao, Yuchen and Zhao, Xiaotao},
  journal={China Economic Review},
  volume={79},
  pages={101951},
  year={2023}
}

@unpublished{warrence2002basics,
  title={Basics of salinity and sodicity effects on soil physical properties},
  author={Warrence, Nikos J and Bauder, James W and Pearson, Krista E},
  year={2002}, 
  note={MSU Extension Water Quality Program}
}

@book{desbureaux2019impact,
  title={The impact of water quality on {GDP} growth: {E}vidence from around the world},
  author={Desbureaux, Sebastien and Damania, Richard and Rodella, Aude-Sophie and Russ, Jason and Zaveri, Esha},
  year={2019},
  publisher={World Bank, Washington, DC}
}

@article{mahmood2014human,
  title={{Human health risk assessment of heavy metals via consumption of contaminated vegetables collected from different irrigation sources in Lahore, Pakistan}},
  author={Mahmood, Adeel and Malik, Riffat Naseem},
  journal={Arabian Journal of Chemistry},
  volume={7},
  number={1},
  pages={91--99},
  year={2014}
}

@online{wto2024,
    author = {{World Trade Organization}},
    title = {{Evolution of trade under the WTO: {H}andy statistics}},
    url = "https://www.wto.org/english/res_e/statis_e/trade_evolution_e/evolution_trade_wto_e.htm#:~:text=World%20trade%20values%20today%20have,the%20WTO%20was%20first%20established.",
    addendum = "(accessed: 22.04.2024)",
    year={2024}
}

@online{aqli2023,
    author = {{AQLI}},
    title = {{India Fact Sheet}},
    url = "https://aqli.epic.uchicago.edu/wp-content/uploads/2023/08/India-FactSheet-2023_Final.pdf",
    year={2023}
}

@book{banga2012twenty,
  title={{Twenty years of India's liberalization: Experiences and lessons}},
  author={Banga, Rashmi and Das, Abhijit},
  year={2012},
  publisher={United Nations Publications}
}

@article{solon2015we,
  title={{What are we weighting for?}},
  author={Solon, Gary and Haider, Steven J and Wooldridge, Jeffrey M},
  journal={Journal of Human Resources},
  volume={50},
  number={2},
  pages={301--316},
  year={2015}
}

@article{chakraborty2014estimation,
  title={{Estimation of Water Pollution Content in India’s Foreign Trade}},
  author={Chakraborty, Debesh and Mukhopadhyay, Kakali and Chakraborty, Debesh and Mukhopadhyay, Kakali},
  journal={Water Pollution and Abatement Policy in India: A Study from an Economic Perspective},
  pages={119--140},
  year={2014},
  publisher={Springer}
}

@article{barrows2021foreign,
  title={{Foreign demand, developing country exports, and CO2 emissions: Firm-level evidence from India}},
  author={Barrows, Geoffrey and Ollivier, H{\'e}l{\`e}ne},
  journal={Journal of Development Economics},
  volume={149},
  pages={102587},
  year={2021},
  publisher={Elsevier}
}

@article{goyal1996political,
  title={{Political economy of India’s economic reforms}},
  author={Goyal, Surinder K},
  journal={Institute for Studies in Industrial Development (ISID) Working Paper},
  volume={4},
  year={1996}
}

@incollection{copeland2022globalization,
  title={{Globalization and the environment}},
  author={Copeland, Brian R and Shapiro, Joseph S and Taylor, M Scott},
  booktitle={{Handbook of International Economics}},
  volume={5},
  pages={61--146},
  year={2022},
  publisher={Elsevier}
}

@article{anukriti2019women,
  title={{Women’s worth: Trade, female income, and fertility in India}},
  author={Anukriti, Sharma and Kumler, Todd J},
  journal={Economic Development and Cultural Change},
  volume={67},
  number={3},
  pages={687--724},
  year={2019},
  publisher={The University of Chicago Press Chicago, IL}
}
